\documentclass[showpacs,aps,prd,floatfix,amsmath,amssymb,nofootinbib,superscriptaddress]{revtex4-2}
\usepackage{graphicx}
\usepackage{amssymb}
\usepackage{bbm}
\usepackage{hyperref}
\usepackage{slashed}
\usepackage{float}
\usepackage{xcolor}
\usepackage{subfigure}
\usepackage{dcolumn}
\usepackage{hyperref}

\usepackage{multirow}
\begin{document}

\title{$CP$ violation in the $HZZ$ vertex and left-right asymmetries}
\author{A. I. Hern\'andez-Ju\'arez}
\email{alan.hernandez@cuautitlan.unam.mx}
\affiliation{Departamento de F\'isica, FES-Cuautitl\'an, Universidad Nacional Aut\'onoma de M\'exico, C.P. 54770, Estado de M\'exico, M\'exico.}
\author{R. Gait\'an}
\affiliation{Departamento de F\'isica, FES-Cuautitl\'an, Universidad Nacional Aut\'onoma de M\'exico, C.P. 54770, Estado de M\'exico, M\'exico.}
\date{\today}

\date{\today}

\begin{abstract}
We investigate new contributions to the $HZZ$ vertex from the Flavor Changing Neutral Current (FCNC) involving the Higgs and $Z$ bosons. Our calculations reveal that the form factors $h_2^V$ and $h_3^V$ ($V=H$, $Z$) can be induced through these couplings, and we present our results in terms of the Passarino-Veltman scalar functions. Using the current limits on $H\overline{t}c$ and $Z\overline{t}c$ couplings, we determine that the new contributions to the $CP$-conserving form factor $h_2^V$ are small compared to the Standard Model (SM) predictions. However, for the $CP$-violating form factor $h_3^V$, the contributions can reach values as high as $10^{-6}$, five orders of magnitude larger than in the SM. Furthermore, we examine how these results influence the left-right asymmetries in the processes $H^\ast\to ZZ$ and $Z^\ast\to ZH$. Our findings suggest that significant deviations from SM predictions may occur when considering FCNC contributions.

\end{abstract}


\date{\today}

\maketitle
\section{ Introduction}
In the SM of particle physics, the Higgs boson, discovered in 2012 at the Large Hadron Collider (LHC) \cite{CMS:2012qbp, ATLAS:2012yve}, is the remanent of the Brout-Englert-Higgs mechanism that gives mass to the gauge bosons and fermions \cite{PhysRevLett.13.321, PhysRevLett.13.585, PhysRevLett.13.508}. Since the Higgs boson discovery, notable progress has been made in measuring its properties \cite{ATLAS:2022vkf, CMS:2022dwd}. Recently, the couplings of the Higgs boson with weak bosons have attracted considerable attention, particularly after the LHC reported that the signal strength of the $H \rightarrow Z\gamma$ decay is twice the predicted value by the SM \cite{CMS:2022ahq, ATLAS:2023yqk}. Furthermore, for the first time, the evidence of a pair of $Z$ bosons produced via an off-shell Higgs boson was announced by the ATLAS and CMS collaborations \cite{CMS:2022ley, ATLAS:2023dnm}. This finding also facilitated the determination of $\Gamma_H$ by analyzing the ratio between the on-shell and off-shell $Z$ pair production rates \cite{Caola:2013yja, Campbell:2013una}. Moreover, the $HZZ$ vertex has been proposed to be sensitive to quantum entanglement effects at the LHC \cite{Aguilar-Saavedra:2022wam, Bernal:2023ruk, Aguilar-Saavedra:2024whi, Bernal:2024xhm, Sullivan:2024wzl}.

 The $HZZ$ vertex with anomalous couplings can be induced through the following effective Lagrangian
\begin{align}
\label{Lag2}
\mathcal{L}=&\frac{g }{c_W}m_Z \left[\frac{(1-a_Z)}{2}  H Z_\mu Z^\mu+\frac{1}{2m^2_Z} \Big\{\hat{b}_Z HZ_{\mu\nu}Z^{\mu\nu}+\hat{c}_Z HZ_\mu\partial_\nu Z^{\mu\nu}+ \tilde{b}_Z H Z_{\mu\nu}\tilde{Z}^{\mu\nu}\Big\}\right],
\end{align}
 where $a_Z$ corresponds to the tree-level correction, while $\hat{b}_Z$ arises at the one-loop level within the SM and is of order $10^{-2}$  \cite{Kniehl:1990mq, Phan:2022amy, Hernandez-Juarez:2023dor}. The anomalous coupling $\hat{c}_Z$ is also expected to emerge at the one-loop level; however, it has not been identified yet in SM calculations. The $a_Z$, $\hat{b}_Z$ and $\hat{c}_Z$ couplings are $CP$-conserving \cite{Kniehl:1990mq}, whereas $\tilde{b}_Z$ is $CP$-violating and could be generated at the three-loop level in the SM, with an approximate magnitude of $10^{-11}$ \cite{Soni:1993jc}.   At the LHC, bounds on the effective fractional cross sections $f_{a_i}$  have been derived from the decays $H\to4\ell$ \cite{CMS:2022ley, CMS:2021nnc} and $H\to\tau\tau$ \cite{CMS:2022uox}. The $f_{a_i}$ approach minimizes uncertainties and is independent of the coupling parametrization \cite{CMS:2017len, CMS:2019ekd}; however, this methodology restricts the determination of limits to the ratios $\hat{c}_Z/\hat{b}_Z$ and $\tilde{b}_Z/\hat{b}_Z$. 
In contrast, by integrating the LHC results in Ref. \cite{CMS:2022ley} with the theoretical calculation of $\hat{b}_Z$, individual constraints on the $\hat{c}_Z$ and $\tilde{b}_Z$ anomalous couplings of order $10^{-2}-10^{-4}$ were established in Ref. \cite{Hernandez-Juarez:2023dor}. Tighter bounds can be achieved using the stringent constraints on the effective ratios $f_{a_i}$ presented in Ref. \cite{CMS:2022uox}, which reports an improvement of approximately 20-50\% in these limits through the combination of results from the decays $H\to4\ell$ and $H\to\tau\tau$. Nevertheless, these bounds are of comparable order of magnitude to those used in  Ref. \cite{Hernandez-Juarez:2023dor}. Therefore, significant changes for the constraints on $\hat{c}_Z$ and $\tilde{b}_Z$ outlined above are not anticipated.  The phenomenology of the $H^\ast ZZ$ coupling at the LHC and future colliders has been extensively studied by numerous authors \cite{Bolognesi:2012mm, Anderson:2013afp, Sahin:2019wew, Gritsan:2020pib, Goncalves:2020vyn, Azatov:2022kbs, Nguyen:2022ubf}. Additionally, the cases $HZZ^\ast$ and $HZ^\ast Z^\ast$ have relevant implications at colliders  \cite{Kniehl:1991hk,Hagiwara:1993sw,Hagiwara:2000tk,Kniehl:2001jy,Biswal:2005fh,Choudhury:2006xe,Godbole:2007cn,Dutta:2008bh,Rindani:2009pb,Cakir:2013bxa,Rao:2019hsp,Kumar:2019bmk, Rao:2020hel, Rao:2021eer,Chen:2020gae,Bizon:2021rww,Sharma:2022epc,Rao:2022olq,Bittar:2022wgb,Gritsan:2022php,Chen:2022mre, Rao:2023jpr, Tran:2022fdb, Sharma:2025ceq, Das:2025ifh}. The $H^\ast\to ZZ$ decay is included in publicly available codes such as \texttt{HDECAY} \cite{Djouadi_1998hd} and \texttt{PROPHECY4F} \cite{PhysRevD.74.013004}.
 \begin{figure}[H]
\begin{center}
\includegraphics[width=10cm]{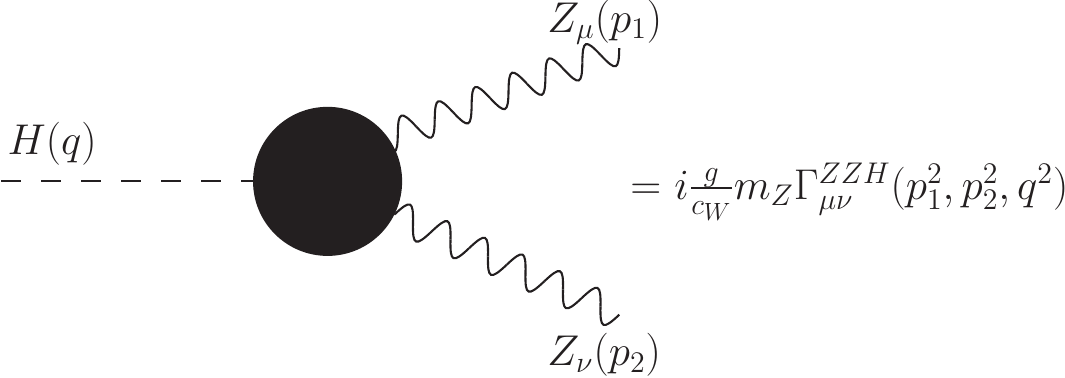}
\caption{Nomenclature for the $HZZ$ coupling and the $\Gamma^{ZZH}_{\mu\nu}$ vertex function.} \label{ZHHvertex}
\end{center}
\end{figure}
 
From Lagrangian \ref{Lag2} and using the nomenclature in Fig \ref{ZHHvertex}, the vertex function can be written as follows
\begin{align}
\label{vertex}
\Gamma_{\mu\nu}^{ZZH}=h^V_1(q^2,p_1^2,p_2^2) g_{\mu\nu}+\frac{h_2^V(q^2,p_1^2,p_2^2)}{m_Z^2} p_{1\nu}p_{2\mu}+\frac{h_3^V(q^2,p_1^2,p_2^2) }{m_Z^2}\epsilon_{\mu\nu\alpha\beta}p_{1}^\alpha p_{2}^\beta,
\end{align}
where $V$ denotes the off-shell boson. Following the kinematics $H^\ast\rightarrow ZZ$ ($Z^\ast\rightarrow ZH$), the form factors $h_i^V$ can be expressed as
\begin{align}
& h^V_1(q^2,p_1^2,p_2^2)=1+ a_Z-   \hat{b}_Z \frac{q^2-p_1^2-p_2^2}{m_Z^2}+\frac{ \hat{c}_Z}{2 m_Z^2}p_1^2+p_2^2, \label{H11}\\
&h^V_2(q^2,p_1^2,p_2^2)=\pm 2 \hat{b}_Z, \label{H22}\\
&h^V_3(q^2,p_1^2,p_2^2)= \pm 2 \widetilde{b}_Z. \label{H33}
\end{align}
For simplicity, we set $a_Z=0$ in this work. Since $\hat{b}_Z$ has an imaginary part in the SM \cite{Hernandez-Juarez:2023dor}, we expect that  $\hat{c}_Z$, $\hat{b}_Z$, $\widetilde{b}_Z$ are complex quantities. The Lagrangian in Eq. \eqref{Lag2} necessitates real anomalous couplings to be Hermitian. However, it uses an effective approach to describe the $HZZ$ interaction, which is valid only at the Born level; thus, the operators that induce the anomalous couplings at higher orders are not included in the $HZZ$ Lagrangian \cite{Hagiwara:1986vm}. These operators do not strictly require real anomalous couplings to maintain a Hermitian Lagrangian.

The imaginary part, together with the $CP$-violating form factor, may give rise to intriguing new physics effects that could be observed through polarized observables at the LHC \cite{CHANG1993286, Ilakovac:1993pt, Grzadkowski:1994qk, Godbole:2007cn, Cao:2020npb, Hernandez-Juarez:2023dor, Hernandez-Juarez:2024zpk}. The phenomenology involving a pair of polarized $Z$ bosons in the process $pp\rightarrow H^\ast\rightarrow ZZ$ at the LHC has been addressed in Refs. \cite{Ballestrero:2019qoy, Maina:2020rgd, Maina:2021xpe, Javurkova:2024bwa, Grossi:2024jae}. The polarizations of gauge bosons are particularly noteworthy at the LHC and are currently under investigation across various processes. For instance, the ATLAS collaboration has reported evidence of longitudinally polarized $W^\pm Z$ and $ZZ$ boson pairs \cite{ATLAS:2023zrv, ATLAS:2022oge}. Additionally, polarization fractions of the $Z$ bosons have been analyzed by the LHCb, ATLAS, and CMS collaborations \cite{LHCb:2022tbc, ATLAS:2016rnf, ATLAS:2023lsr, CMS:2015cyj}. Studies involving gauge boson polarizations at the LHC include $W^{\pm}W^{\pm}$ production \cite{CMS:2020etf}, $W+$Jets events \cite{CMS:2011kaj, ATLAS:2012au}, $W^{\pm}Z$ production \cite{ATLAS:2019bsc}, and the generation of $W$ bosons in $t\bar{t}$ events and top decays \cite{CMS:2016asd, ATLAS:2016fbc, CMS:2020ezf, ATLAS:2022rms}. The potential for producing polarized gauge bosons has been incorporated into event generators like \texttt{MadGraph5\_aMC@NLO} \cite{BuarqueFranzosi:2019boy} and \texttt{SHERPA} \cite{Hoppe:2023uux}. 

In this study, we examine the FCNC contributions of the Higgs and $Z$ bosons to the $HZZ$ vertex, with a particular focus on the $CP$-violating form factor $h_3^V$ ($V=H$, $Z$). Furthermore, we investigate the potential for new left-right asymmetries arising from the polarizations of the $Z$-bosons. The structure of this work is organized as follows:  in Sec. \ref{secFCNCcon}, we calculate the new contributions to the $HZZ$ vertex resulting from FCNC couplings mediated by the $H$ and $Z$ bosons. Next, in Sec. \ref{LRAsym}, we analyze the left-right asymmetries that can be induced by the $CP$-violating form factor $h_3^V$. Finally, in Sec. \ref{senNum}, we present a numerical analysis of our results, with our conclusions summarized in Sec. \ref{concl}.

\section{FCNC contributions to the $HZZ$ coupling}\label{secFCNCcon}

To generate $CP$-violating contributions to the $HZZ$ vertex, we consider the following effective Lagrangian, which induces FCNC couplings mediated by the $Z$ and $H$ bosons 
\begin{align}
\label{Lag}
\mathcal{L}=\frac{g}{c_W}\overline{f}_j\gamma_\mu\Big(g^{ij}_V-g^{ij}_A\gamma^5\Big)f_i Z^\mu-\frac{g}{2 m_W}H \overline{f}_j\Big(g^{ij}_S+g^{ij}_P\gamma^5\Big)f_i+\mathcal{H.C},
\end{align}
where $f_{i,j}$ correspond to the SM fermions, while $g^{ij}_{r}$ ($r=V$, $A$, $S$, $P$) are complex couplings. For FCNC couplings involving light quarks mediated by the Higgs boson, the strongest constraints are derived from $B-\overline{B}$, $K^0-\overline{K}^0$ and $D^0-\overline{D}^0$ oscillations \cite{Harnik:2012pb}. The corresponding limits on the $g^{ij}_{V, A}$ couplings are obtained from decays of $K^0$ and $B$ mesons \cite{Buchalla:2000sk, Mohanta:2005gm, Silverman:1991fi, Buras:1998ed, Giri:2003jj}. Bounds on lepton FCNC mediated by the Higgs and $Z$ bosons have been established through the analysis of dipole moments, meson oscillation, and three-body lepton flavor violating decays \cite{Harnik:2012pb, Mohanta:2010yj}. Moreover, these interactions have been analyzed at the LHC \cite{ATLAS:2014vur,CMS:2015qee,ATLAS:2018sky,CMS:2021rsq,ATLAS:2022uhq}. FCNC interactions involving the top quark have been investigated at the LHC through the decays $t\rightarrow Zq$ and $t\rightarrow Hq$ \cite{ATLAS:2023qzr, ATLAS:2023ujo}. In the SM, the corresponding branching ratios are highly suppressed, with $\mathcal{B}(t\rightarrow Zq)$ being of order $10^{-14}$, and $\mathcal{B}(t\rightarrow Hq)$ one order of magnitude smaller \cite{Aguilar-Saavedra:2004mfd}. However, these results could be significantly enhanced by new physics contributions \cite{Abraham:2000kx, Eilam:2001dh, Aguilar-Saavedra:2002phh, Grossman:2007bd, Aguilar-Saavedra:2010ewj, Aguilar-Saavedra:2013qpa, Gaitan:2017tka, Badziak:2017wxn,Liu:2020kxt,Hou:2020chc,Liu:2020bem, Chen:2022dzc, Chen:2023eof}. Given the large mass of the top quark, we expect the most significant contributions from FCNC interactions involving this particle. Constraints on the couplings $Z\overline{t}q$ and $H\overline{t}q$ ($q=c$, $u$) have been obtained from LHC data \cite{Hernandez-Juarez:2022kjx,Hernandez-Juarez:2024pty}, which can be summarized as follows:
\begin{equation}
\label{bound1}
\big|g^{tc}_V\big|\text{, }\big|g^{tc}_A\big|\leqslant 0.0095,
\end{equation}
\begin{equation}
\label{bound2}
\big|g^{tc}_S\big|\text{, }\big|g^{tc}_P\big|\leqslant 0.25 \text{ GeV},
\end{equation}
while the corresponding limits for the up quark are smaller than those presented above, but of the same order of magnitude. 

In the rest of this section, we will calculate the one-loop FCNC contributions to the  $h_2^V$ and $h_3^V$ from factors arising from Lagrangian \eqref{Lag}. To our knowledge, these contributions have not been previously reported. We will consider scenarios that involve an off-shell Higgs boson or an off-shell $Z$ boson, with the following kinematics $H^\ast\rightarrow ZZ$ and $Z^\ast\rightarrow ZH$.

\subsection{Analytical results}

\begin{figure}[H]
\begin{center}
\includegraphics[width=8cm]{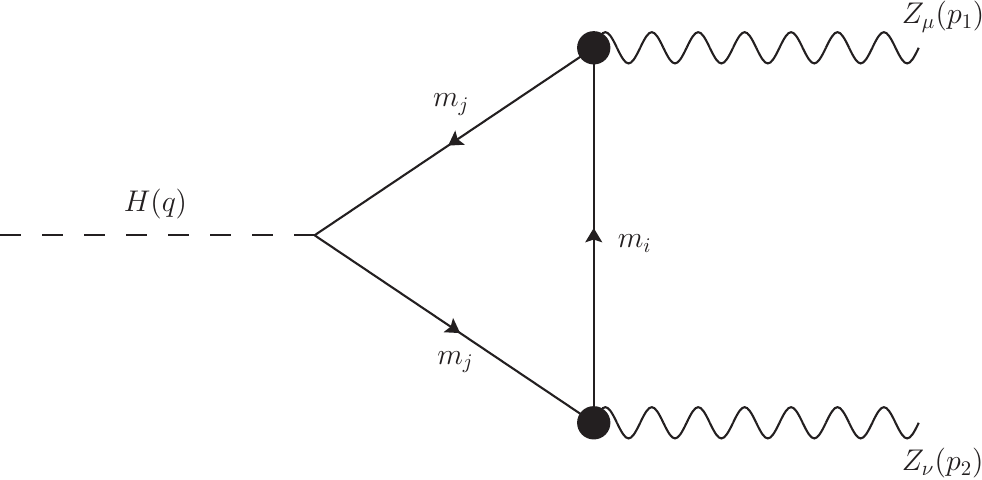}
\caption{One-loop contributions of type I, where only FCNC couplings of the $Z$ boson are involved.} \label{diag1}
\end{center}
\end{figure}

\begin{figure}[H]
\begin{center}
\subfigure{\includegraphics[width=7.5cm]{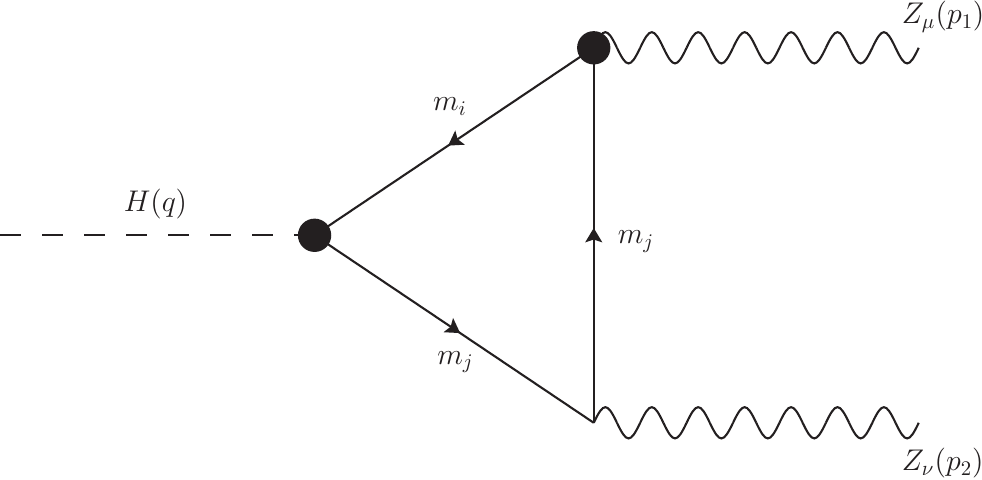}}
\subfigure{\includegraphics[width=7.5cm]{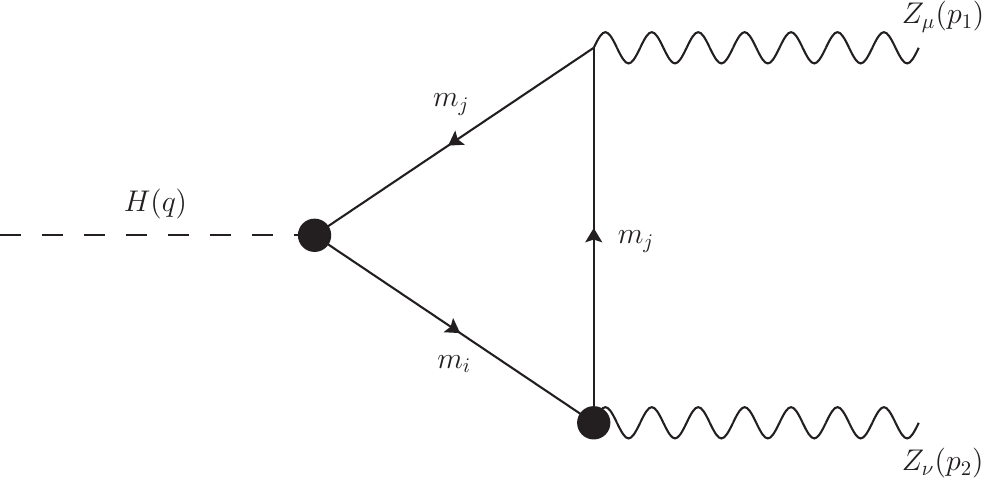}}
\caption{One-loop contributions of type II, where FCNC couplings of the $Z$ and $H$ bosons are involved.} \label{diag2}
\end{center}
\end{figure}

 Two distinct contributions to the $HZZ$ vertex can arise from the FCNC couplings in Lagrangian \eqref{Lag}. The first category of diagrams (Type I), as illustrated in Fig. \ref{diag1}, involves only flavor-violating couplings mediated by the $Z$ boson. The second category (Type II) also includes FCNC couplings of the Higgs boson and is depicted in Fig. \ref{diag2}. For our calculations, we consider additional diagrams arising from both $p_1^\mu \leftrightarrow p_2^\nu$ and $m_i \leftrightarrow m_j$ exchanges. Our results were obtained using the FeynCalc package \cite{Mertig:1990an, Shtabovenko:2016sxi, Shtabovenko:2020gxv, Shtabovenko:2023idz} and are expressed in terms of the Passarino-Veltman scalar functions.

\subsubsection{Diagrams type I}

  By considering all the contributing Feynman diagrams from Fig. \ref{diag1}, we obtained the FCNC contribution to the $CP$-conserving form factor $h_2^V$, which can be expressed as
\begin{align}
\label{h2type1}
h^V_2(I)= -\frac{g^2 m_Z N_f }{4 \pi ^2  c_W
   m_W }\Bigg\{\big|
   g_V^{ij}\big|
   {}^2A_\mathcal{V}^V(K^2,m_i^2,m_j^2)+\big|
   g_A^{ij}\big|
   {}^2A^V_A(K^2,m_i^2,m_j^2)\Bigg\}, \quad V=H\text{, }Z, \quad K=Q\text{, }P,
\end{align}
where $N_f$ corresponds to the number of colors. The functions $A_{\mathcal{V}, A}^V$ ($V=H$, $Z$) can be found in Appendix \ref{Apptype1}. For the off-shell $H$ case, $A_{\mathcal{V}, A}^H$ depend on $Q$, which is defined as $Q\equiv \| q \|$. In the case of the off-shell $Z$ boson, the functions $A_{\mathcal{V}, A}^Z$ depend on $P$, with $P=\|p_1\|$. When $m_j=m_i$, the form factor $h^V_2$ ($V=H$, $Z$) reduces to twice the SM results \cite{Hernandez-Juarez:2023dor}, due to the inclusion of double diagrams corresponding to the \(m_i \leftrightarrow m_j\) exchange. 

For the $CP$-violating form factor $h_3^V$, the FCNC contribution of type I is given as follows
\begin{align}\label{h3type1}
h_3^V(I)=-\frac{ g^2 m_i m_j m_Z N_f}
   {2 \pi ^2 c_W m_W }    {\rm Im}\Big[g_V^{ij}g_A^{ij\ast}\Big]\widetilde{\mathcal{F}}^V(K^2,m_i^2,m_j^2), \quad V=H\text{, }Z, \quad K=Q\text{, }P,
   \end{align}
where the functions $\widetilde{\mathcal{F}}^V$ ($V=H$, $Z$) can also be found in Appendix \ref{Apptype1}.  Notably, at least one complex coupling in the $Z\overline{f}_j f_i$ vertex is required to induce $CP$ violation in the $HZZ$ vertex. For the flavor-conserving scenario $m_j=m_i$, the form factor $h_3^V$ vanishes. Additionally, we anticipate negligible contributions for FCNC involving only light fermions, as $h_3^V$ is proportional to the product $m_i m_j$.

The $A_{\mathcal{V}, A}^V$ and $\widetilde{\mathcal{F}}^V$ ($V=H$, $Z$)  functions presented in Eqs.\eqref{h2type1} and \eqref{h3type1} are free of divergences.

\subsubsection{Diagrams Type II}

From diagrams of type II in Fig. \ref{diag2}, the resulting contribution to the $CP$-conserving form factor can be expressed as
\begin{align}
\label{h2type2}
h^V_2(II)= -\frac{ g^2 m_Z N_f}{2 \pi ^2  c_W m_W }&\Bigg\{ g_V{\rm Re}\Big[g_V^{ij}g_S^\ast\Big]R^V_{1}(K^2,m_i,m_j)+g_A{\rm Re}\Big[g_V^{ij}g_P^\ast\Big]R^V_{2}(K^2,m_i,m_j)\nonumber\\
&+g_A{\rm Re}\Big[g_A^{ij}g_S^\ast\Big]R^V_{3}(K^2,m_i,m_j)+g_V{\rm Re}\Big[g_A^{ij}g_P^\ast\Big]R^V_{4}(K^2,m_i,m_j)\Bigg\},  \quad V=H\text{, }Z, \quad K=Q\text{, }P,
\end{align}
where $g_{V}$ and $g_A$ correspond to the SM vector and axial couplings of the $Z$ boson with fermions.  The functions $R^V_i$ ($i=$1, 2, 3, 4) are presented in Appendix \ref{Apptype2}. For the case where $m_j=m_i$, $g_S=m_i$, and $g_P=0$, our expression simplifies to four times the SM contribution due to the additional Feynman diagrams from the $m_i\leftrightarrow m_j$ and $Z\overline{f}_jf_i$ exchanges.

  The contribution to the form factor associated with $CP$ violation can be expressed as follows:
\begin{align}\label{h3type2}
h_3^V(II)=&\frac{ g^2 m_Z N_f}{2 \pi ^2  c_W m_W }   \Bigg\{  g_A{\rm Im}\Big[g_V^{ij}g_S^\ast\Big]T^V_{1}(K^2,m_i,m_j)+g_V{\rm Im}\Big[g_V^{ij}g_P^\ast\Big]T^V_{2}(K^2,m_i,m_j)\nonumber\\
&+g_V{\rm Im}\Big[g_A^{ij}g_S^\ast\Big]T^V_{3}(K^2,m_i,m_j)+g_A{\rm Im}\Big[g_A^{ij}g_P^\ast\Big]T^V_{4}(K^2,m_i,m_j)\Bigg\}, \quad V=H\text{, }Z, \quad K=Q\text{, }P.
   \end{align}
The functions $T^V_i$ ($i$=1, 2, 3, 4) in Eq. \eqref{h3type2} are also shown in Appendix \ref{Apptype2}. Notably, the pseudoscalar coupling is not necessary to induce $CP$ violation. For the flavor-conserving scenario ($m_j=m_i$), and considering $g_P=0$, the form factor $h_3^V$ vanishes. However, the functions $T^V_{2\text{, }4}$, which are proportional to terms involving the pseudoscalar coupling $g_P$, are not zero when $m_j=m_i$. Therefore, generating \(CP\) violation without FCNC contributions remains feasible if we include a pseudoscalar coupling.
The $R^V_i$ and $T^V_i$ ($i$=1,2,3, 4) functions in  Eqs. \eqref{h2type2} and \eqref{h3type2} are free of divergences.

The new contributions to the form factor $h_3^H$ can lead to left-right asymmetries in the $H^\ast\to ZZ$ process \cite{Hernandez-Juarez:2023dor, Hernandez-Juarez:2024zpk}. Furthermore, the FCNC contributions to $h_2^V$ and $h_3^V$ ($V=H$, $Z$) of types I and II have not been reported previously.

\section{Left-right asymmetries of the $HZZ$ vertex}\label{LRAsym}

The polarized observables of the $HZZ$ vertex have been explored in multiple contexts \cite{Godbole:2007cn, Cao:2020npb, Hernandez-Juarez:2023dor, Hernandez-Juarez:2024zpk}, where the left-right asymmetry $\mathcal{A}^H_{LR}$ for the $H^\ast \rightarrow ZZ$ process is defined as follows:
\begin{equation}
\mathcal{A}^H_{LR}=\frac{\Gamma^L_{H^\ast\rightarrow Z_L Z_L}-\Gamma^R_{H^\ast\rightarrow Z_R Z_R}}{\Gamma^L_{H^\ast\rightarrow Z_L Z_L}+\Gamma^R_{H^\ast\rightarrow Z_R Z_R}}.
\end{equation}
By considering the form factors $h_i^H$ as complex, the $\mathcal{A}^H_{LR}$ asymmetry has been computed in Ref. \cite{Hernandez-Juarez:2023dor} in terms of the real and imaginary parts of $h_1^H$ and $h_3^H$ as follows
\begin{equation}
\label{AsymmetryH}
\mathcal{A}^H_{LR}= 4 m_Z^2 \mathcal{K}(Q)\frac{{\rm Re}\big[h_1^H\big]{\rm Im}\big[h_3^H\big]-{\rm Re}\big[h_3^H\big]{\rm Im}\big[h_1^H\big]}{\mathcal{K}^2(Q)\Big\{{\rm Re}\big[h_3^H\big]^2+{\rm Im}\big[h_3^H\big]^2\Big\}+4m_Z^4 \Big\{{\rm Re}\big[h_1^H\big]^2+{\rm Im}\big[h_1^H\big]^2\Big\}},
\end{equation}
with
\begin{equation}
\mathcal{K}(Q)=\sqrt{Q^2(Q^2-4m_Z^2)}.
\end{equation}
Motivated by this result, we expect that a non-zero left-right asymmetry can also arise in the process $Z^\ast \rightarrow ZH$. With this aim, we define $\mathcal{A}^Z_{LR}$  as
\begin{equation}\label{LRZas}
\mathcal{A}^Z_{LR}=\frac{\Gamma^L_{Z^\ast\rightarrow Z_L H}-\Gamma^R_{Z^\ast\rightarrow Z_R H}}{\Gamma^L_{Z^\ast\rightarrow Z_L H}+\Gamma^R_{Z^\ast\rightarrow Z_R H}}.
\end{equation}
To obtain the analytic expression for the $\mathcal{A}^{Z}_{LR}$ asymmetry, we begin by calculating the polarized width decays $\Gamma^\lambda_{Z^\ast\rightarrow Z_\lambda H}$ under the following kinematic conditions: the process $Z^\ast(p_1)\rightarrow Z(p_2)H(q)$ occurs in the rest frame of the $Z^\ast(p_1)$ boson, with the $Z(p_2)$ moving along the $x$-axis. In this context, the polarization vectors of the $Z(p_2)$ boson are:
\begin{align}
\label{polaVec}
   & \epsilon(0)=\frac{1}{2m_Z P}\Big(\sqrt{(P^2+m_Z^2-m_H^2)^2-4m_Z^2 P^2}, P^2+m_Z^2-m_H^2, 0, 0 \Big),   \\
    &  \epsilon(R/L)=\frac{1}{\sqrt{2}}\Big(0,0,-i,\pm 1 \Big).
\end{align}
Then, the polarized width decays can be expressed as 
\begin{equation}\label{widtpolgen}
\Gamma^\lambda_{Z^\ast\rightarrow Z_\lambda H}=\frac{\sqrt{(P^2+m_Z^2-m_H^2)^2-4m_Z^2 P^2}}{16\pi P^3}\mathcal{M}^2(\lambda)\text{,}\quad \lambda=L\text{, }R\text{, }0.
\end{equation}
where $\lambda$ denotes the polarization of the on-shell $Z$ boson and $\mathcal{M}^2(\lambda)$ is the squared polarized amplitude. In Eq \eqref{widtpolgen}, we do not average over the initial polarizations because the off-shell $Z(p_1)$ boson corresponds to a propagator in a collider process.

 The right and left polarized amplitudes are given as
\begin{align}
\label{LRamplitude}
\mathcal{M}^2(R/L)=&-\frac{g^2}{4 c_W^2
   m_Z^4} \Bigg\{ m_Z^2
   \bigg[ -4
   m_Z^4
   \left({\rm Re}\big[h_1^Z\big]{}^2+
   {\rm Im}\big[h_1^Z\big]{}^2\right)+\big({\rm Re}\big[h_3^Z\big]{}^2+{\rm Im}\big[h_3^Z\big]{}^2\big)
   \big(2 m_H^2
   \left(m_Z^2+P^2\right)-m_H^4\nonumber\\
   &-\left(m_Z^2-P^2\right){}^2\big)
   \bigg]\pm 4 m_Z^4
 \left(   {\rm Re}\big[h_1^Z\big]{\rm Im}\big[h_3^Z\big]-{\rm Im}\big[h_1^Z\big]
   {\rm Re}\big[h_3^Z\big]\right)
   \sqrt{(P^2+m_Z^2-m_H^2)^2-4m_Z^2 P^2}\Bigg\}.
\end{align}
We observe that analogous to the $H^\ast\to ZZ$ process, the amplitudes for transversely polarized states only exhibit a dependence on the $h_1^Z$ and $h_3^Z$ form factors \cite{Hernandez-Juarez:2023dor}. 

To ensure a comprehensive analysis, we have computed the amplitude corresponding to longitudinal polarization:
\begin{align}
\label{0amp}
\mathcal{M}^2(0)=&\frac{g^2}{16
   c_W^2 m_Z^6}
   \Bigg\{4
   m_Z^4 \bigg({\rm Re}\big[h_1^Z\big]{}^2+{\rm Im}\big[h_1^Z\big]{}^2\bigg)
   \left(-2 m_H^2
   \left(m_Z^2+P^2\right)+m_H^4-2
   P^2 m_Z^2+5 m_Z^4+P^4\right)\nonumber\\
   &+\bigg[{\rm Re}\big[h_2^Z\big]{}^2+{\rm Im}\big[h_2^Z\big]{}^2\bigg] \big[(m_H-m_Z)^2-P^2\big]
   \big[(m_H+m_Z)^2-P^2\big]
   \big[m_H^2-3 m_Z^2-P^2\big]
   \big[m_H^2+m_Z^2-P^2\big]
   \nonumber\\
   &
   +4
   m_Z^2 \bigg({\rm Re}\big[h_1^Z\big]
   {\rm Re}\big[h_2^Z\big]+{\rm Im}\big[h_1^Z\big]
   {\rm Im}\big[h_2^Z\big]\bigg)
   \left(-m_H^2+m_Z^2+P^2\right)
   \left(-2 m_H^2
   \left(m_Z^2+P^2\right)+m_H^4+\left(m_Z^2-P^2\right){}^2\right)\Bigg\}.
\end{align}
The absence of the $CP$-violating form factor $h_3^Z$ in Eq. \eqref{0amp} suggests that the longitudinal polarization of the $Z$ boson does not offer an opportunity for detecting $CP$-violating effects. On the other hand, we note a difference in sign between the left- and right-polarized amplitudes in the last term of Eq. \eqref{LRamplitude}. This finding leads to a non-zero $\mathcal{A}^Z_{LR}$ asymmetry, which is sensitive to $CP$ violation and can be expressed as follows
\begin{equation}
\label{AsymmetryZ}
\mathcal{A}^Z_{LR}=4  m_Z^2\ \mathcal{K}(P) \frac{
{\rm Im}\big[h_3^Z\big]
   {\rm Re}\big[h_1^Z\big]-{\rm Im}\big[h_1^Z\big]
   {\rm Re}\big[h_3^Z\big]
  }{
  \mathcal{K}^2(P) \Big\{ {\rm Re}\big[h_3^Z\big]{}^2+{\rm Im}\big[h_3^Z\big]{}^2\Big\}+4
   m_Z^4
   \Big\{{\rm Im}\big[h_1^Z\big]{}^2+
   {\rm Re}\big[h_1^Z\big]{}^2\Big\}
   },
\end{equation}
where the $\mathcal{K}(P)$ function is given by
\begin{equation}
\mathcal{K}(P)=  \sqrt{(P^2+m_Z^2-m_H^2)^2-4m_Z^2 P^2}.
\end{equation}
The $\mathcal{A}^Z_{LR}$ asymmetry exhibits a structure similar to that reported in the $H^\ast\to ZZ$ case. Additionally, we observe that to achieve non-vanishing $\mathcal{A}_{LR}^V$ ($V=H$, $Z$) asymmetries, $CP$-violating and complex anomalous couplings are necessary. In the SM, the $h_1^V$ ($V=H$, $Z$) form factor is known to be complex \cite{Hernandez-Juarez:2023dor}. Therefore, a non-zero $h_3^V$ ($V=H$, $Z$) would give rise to the left-right asymmetries discussed in this section. This work presents the $\mathcal{A}^Z_{LR}$ asymmetry for the first time.

\section{Numerical analysis}\label{senNum}

We will now assess the FCNC contributions to the $h_2^V$ and $h_3^V$ ($V=H$, $Z$) form factors, as well as the $\mathcal{A}^V_{LR}$ asymmetries. Given that we anticipate significant results from FCNC couplings involving the top quark, we will consider the bounds outlined in Eqs. \eqref{bound1} and \eqref{bound2}. Furthermore, our analysis will focus exclusively on energy regions where the $ZZ$ and $HZ$ pairs can be produced on-shell. To numerically evaluate the Passarino-Veltman scalar functions, we utilized the LoopTools package \cite{Hahn:1998yk}.

\subsection{Contributions of Type I}
We begin by analyzing the contributions of type I for both scenarios: the $H^\ast ZZ$ and $Z^\ast ZH$ vertex. To maximize the contributions to the $CP$-conserving form factor, we will use the upper bounds on the norms of the $g^{tc}_V$ and $g^{tc}_A$ couplings in Eq. \eqref{bound1}. 

For the $CP$-violating form factor, it is useful to define
\begin{equation}
\hat{h}_3^V(I)=\frac{h_3^V(I)}{{\rm Im}\Big[g_V^{ij}g_A^{ij\ast}\Big]},\quad V=H\text{, }Z,
\end{equation}
which allows us to analyze its contributions in a model-independent manner.

\subsubsection{$H^\ast ZZ$ vertex}

\begin{figure}[H]
\begin{center}
\subfigure{\includegraphics[width=9cm]{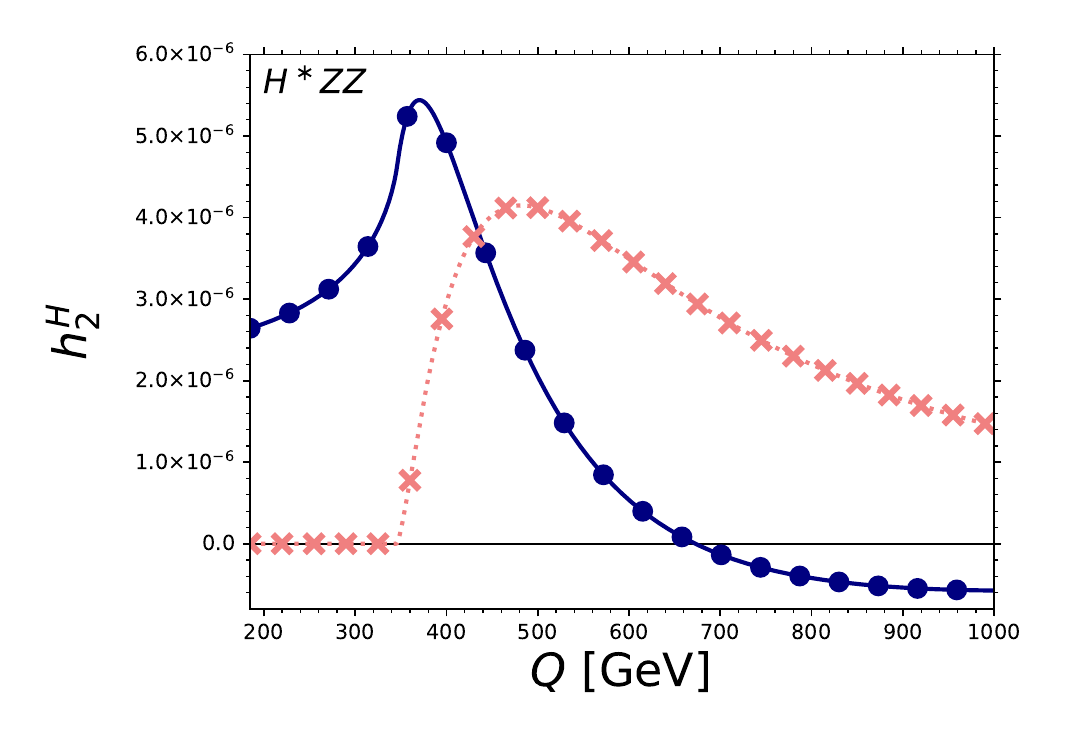}\label{H*ZZplot1a}}\hspace{-.55cm}
\subfigure{\includegraphics[width=9cm]{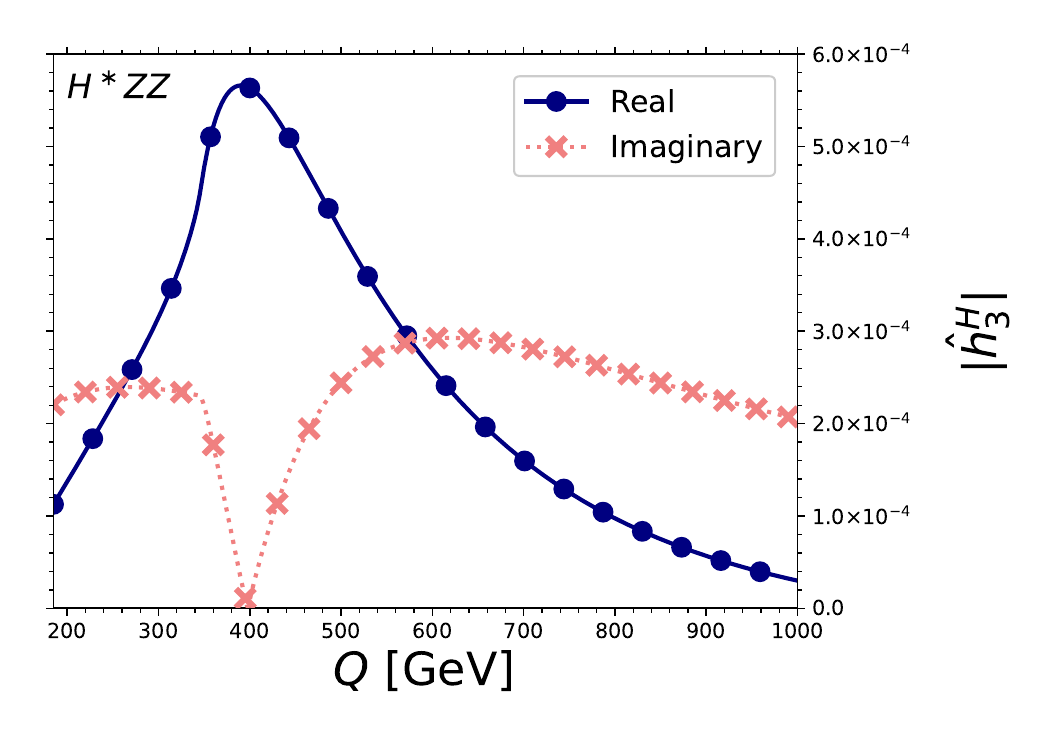}}\hspace{-.01cm}
\caption{The FCNC contributions of the type I to the form factors $h_2^H$ (left plot) and $\hat{h}_3^H$ (right plot) as a function of $Q$. We only consider the $Z\overline{t}c$ coupling.} \label{H*ZZplot1}
\end{center}
\end{figure}

In Fig. \ref{H*ZZplot1}, we present the behavior of the real and imaginary parts of $h^H_2(I)$ (left plot) and $\hat{h}^H_3(I)$ (right plot) as a function of $Q$. Our analysis focuses solely on the $Z\overline{t}c$ contributions, while the contributions from lighter fermions will be addressed later. The absorptive parts emerge when the particles in the loop that couple to the $V^\ast$ boson can be on-shell \cite{Cutkosky:1960sp, Hernandez-Juarez:2021xhy}. For the $CP$ conserving form factor $h^H_2$, we find that its contributions can reach values up to order $10^{-6}$, which is four and three orders of magnitude smaller than those predicted in the SM and the current limits \cite{Hernandez-Juarez:2023dor}, respectively. The real and imaginary parts exhibit a decreasing trend at high energies. Notably, we observe that below the threshold energy of $Q=2m_t$, the absorptive part of $h_2^H$ is approximately of order $10^{-10}$. The contributions to the imaginary part at these lower energy levels are exclusively attributed to the diagrams depicted in Figure \ref{diag1}, wherein the Higgs boson interacts with a $\overline{c}c$ pair. Beyond this threshold energy, additional diagrams involving the $H\overline{t}t$ coupling contribute significantly. As a result, the magnitude of ${\rm Im}\big[h_2^H \big]$ is enhanced and becomes comparable to that of the real part. Remarkably, around $Q\approx 500$ GeV, the absorptive part exceeds the magnitude of ${\rm Re}\big[h_2^H \big]$. 

For the $CP$-violating form factor, we find that the contributions to its real and absorptive parts are of order $10^{-4}$. In contrast to the form factor $h_2^H$, the imaginary part of the $CP$-violating form factor exhibits significantly larger contributions for energies $Q\leqslant 2 m_t$. Furthermore, its magnitude dominates at both low and high energy levels. However, at energy levels around 400 GeV, the real part is the main contribution. From the limits in Eq. \eqref{bound1}, we estimate that ${\rm Im}\big[g_V g^\ast_A \big]$ can reach values of order $10^{-5}$. In this context, the real and imaginary parts of $h_3^H$ could achieve magnitudes between $10^{-8}$ and $10^{-9}$. While this result is quite small compared to the current constraints on $h_3^H$ \cite{Hernandez-Juarez:2023dor}, it is three orders of magnitude larger than the SM prediction \cite{Soni:1993jc}.

For the light fermion FCNC contributions, we examine an optimistic scenario where their couplings are comparable to the values used for the $Z\overline{t}c$ interaction, i.e., they are of order $10^{-3}$. By neglecting the mass of the light quarks to avoid non-perturbative effects, we find that the contributions from the $Z\overline{t}u$ and $Z\overline{c}u$ couplings to the form factor $h_2^H(I)$ are two and six orders of magnitude smaller than in Fig. \eqref{H*ZZplot1}, respectively. For the contributions from FCNC involving down-type quarks and leptons, we estimate values of orders $10^{-10}-10^{-12}$. However, the constraints on these FCNC couplings are tighter than the values considered within this analysis \cite{Harnik:2012pb, Giri:2003jj, Mohanta:2010yj, Chen:2023eof}. As a result, their contributions to $h_2^H(I)$ will be significantly smaller than those outlined above. 

Regarding the $CP$-violating form factor, we find that FCNC contributions from light fermions are negligible compared with those obtained in Fig. \eqref{H*ZZplot1}, since $h_3^V\sim m_im_j$ in Eq. \eqref{h3type1}. 
We observe a similar pattern for the contributions of light fermions to the $Z^\ast ZH$ vertex and type II diagrams. Therefore, these contributions will not be examined further in this work.



\subsubsection{$Z^\ast ZH$ vertex}

\begin{figure}[H]
\begin{center}
\subfigure{\includegraphics[width=9cm]{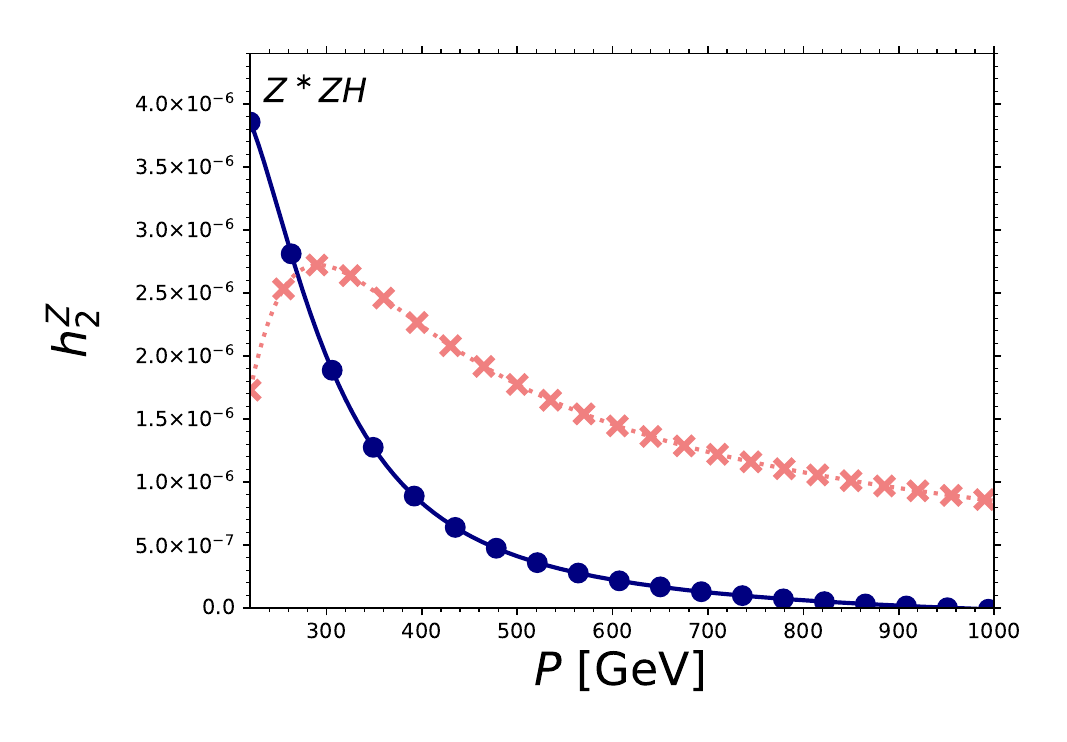}}\hspace{-.55cm}
\subfigure{\includegraphics[width=9cm]{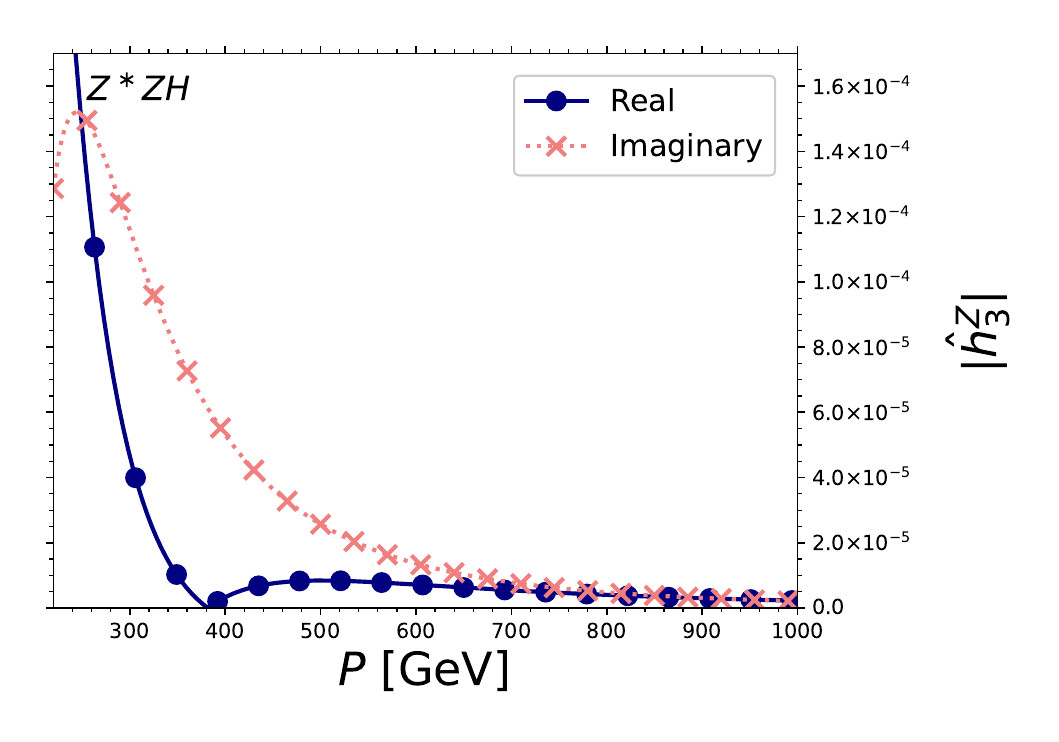}}\hspace{-.01cm}
\caption{The FCNC contributions of the type I to the form factors $h_2^Z$ (left plot) and $\hat{h}_3^Z$ (right plot) as a function of $P$. We only consider the $Z\overline{t}c$ coupling.} \label{Z*ZHplot1}
\end{center}
\end{figure}

We now show the behavior of $h^Z_2(I)$ (left plot) and $\hat{h}^Z_3(I)$ (right plot) as a function of $P$, in Fig. \ref{Z*ZHplot1}. For $h_2^Z$, we observe a pattern similar to the previous case. Both the real and imaginary parts reach magnitudes of order $10^{-6}$, with the absorptive part dominating at high energies. However, for values of $P$ less than $2m_t$, the magnitudes of both the imaginary and real parts are comparable, contrasting with the behavior observed for the scenario involving an off-shell Higgs boson. In the case of the $Z^\ast ZH$ vertex, the amplitudes of the contributing diagrams behave differently, as all exhibit an imaginary component. This phenomenon arises because the $\overline{t}c$ and $\overline{c}t$ pairs, which couple to the $Z^\ast$ gauge boson in Fig \ref{diag1}, can be on-shell for energies $P<2m_t$. Therefore, in the $Z^\ast ZH$ vertex, the absorptive part is not negligible at low values of $P$. 

 For the $CP$-violating form factor  $\hat{h}^Z_3(I)$,  both the real and imaginary parts can be as large as those found in the $H^\ast ZZ$ scenario, although they are significantly smaller as $P$ increases. The imaginary part becomes the dominant contribution in the energy range 300 GeV $\lesssim P\lesssim$ 700 GeV. At high values of $P$, the magnitudes of the real and absorptive parts become similar but of order $10^{6}$. For ${\rm Im}\big[g_V g^\ast_A \big]\sim10^{-5}$, the form factor $h_3^Z$ can achieve values of order $10^{-8}$, which is three orders of magnitude larger than the SM prediction. 
 

\subsection{Contributions of Type II}\label{subtypeIInum}

The form factors obtained from diagrams of type II require a different approach, as they can be expressed through four combinations of the $g_r$ ($r=V$, $A$, $S$, $P$) couplings. To assess the FCNC contributions numerically, we will examine four scenarios that align with the bounds outlined in Eqs. \eqref{bound1} and \eqref{bound2}.

\begin{itemize}
  \item \textbf{Scenario $I$}: As the pseudoscalar coupling in Lagrangian \eqref{Lag} is not essential for inducing $CP$ violation, we focus on the case where $g_P=0$. We employ the upper limits from Eqs. \eqref{bound1}-\eqref{bound2} to determine the values of the real and imaginary parts of the various combinations involved in the form factors $h_2^V(II)$ and $h_3^{V}(II)$. The resulting values are 
  \begin{align}
    & {\rm Re}\big[g_{r}g_S^\ast\big] =0.00237\ \text{GeV},  \quad r=V\text{, }A,  \\
    &  {\rm Im}\big[g_{r}g_S^\ast\big]=0.00237\ \text{GeV},  \quad r=V\text{, }A.
\end{align}
  \item \textbf{Scenario $II$}: In this scenario, we analyze the impact of negative couplings on the behavior of the form factors. The real and imaginary parts that include the pseudoscalar coupling will be negative. Furthermore, we will consider the magnitudes of the couplings to be half of the values used in the previous case.
   \begin{align}
    & {\rm Re}\big[g_{r}g_S^\ast\big]=-{\rm Re}\big[g_{r}g_P^\ast\big] =0.0011\ \text{GeV}, \quad r=V\text{, }A, \\
    &  {\rm Im}\big[g_{r}g_S^\ast\big]=-{\rm Im}\big[g_{r}g_P^\ast\big] =0.0011\ \text{GeV},  \quad r=V\text{, }A.
\end{align}
  \item \textbf{Scenario $III$}: Similar to the previous case, but with different negative values of the real and imaginary parts:
   \begin{align}
    & {\rm Re}\big[g_{V}g_S^\ast\big]={\rm Re}\big[g_{A}g_P^\ast\big]=-{\rm Re}\big[g_{A}g_S^\ast\big]=-{\rm Re}\big[g_{V}g_P^\ast\big] =0.0011\ \text{GeV}, \\
    & {\rm Im}\big[g_{V}g_S^\ast\big]={\rm Im}\big[g_{A}g_P^\ast\big]=-{\rm Im}\big[g_{A}g_S^\ast\big]=-{\rm Im}\big[g_{V}g_P^\ast\big] =0.0011\ \text{GeV}.
\end{align}
  \item \textbf{Scenario $IV$}: In this scenario, we only consider the contributions from the pseudoscalar coupling ($g_S=0$). Furthermore, as the $CP$-violating form factor $h_3^V$ is not vanishing for the flavor-conserving case, we analyze the contributions to $h_3^V$ that involve only top quarks in the loop. For the $h_2^V$ ($V=H$, $Z$) form factor, these contributions are of order $10^{-20}$ and can be neglected. We utilize the same values as those considered in scenario $I$:
  \begin{align}
    & {\rm Re}\big[g_{r}g_P^\ast\big] =0.00237\ \text{GeV},  \quad r=V\text{, }A,  \\
    &  {\rm Im}\big[g_{r}g_P^\ast\big]=0.00237\ \text{GeV},  \quad r=V\text{, }A.
\end{align}

\end{itemize}

\subsubsection{$H^\ast ZZ$}

\begin{figure}[H]
\begin{center}
\subfigure{\includegraphics[width=9cm]{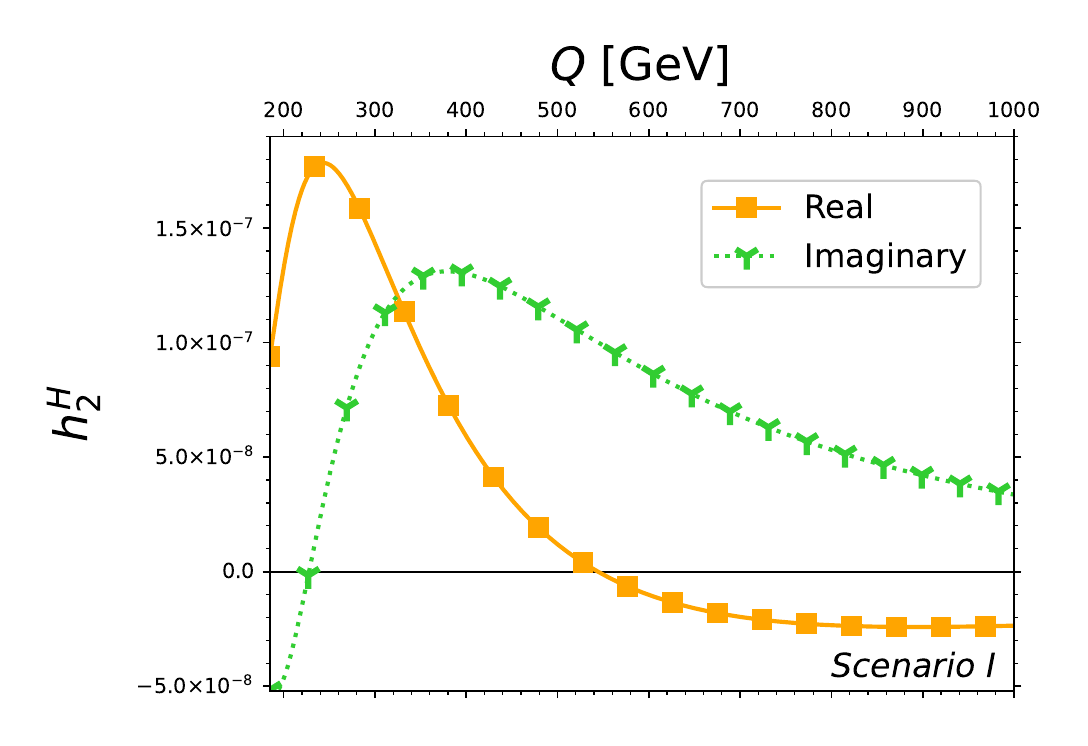}}\hspace{-.58cm}
\subfigure{\includegraphics[width=9cm]{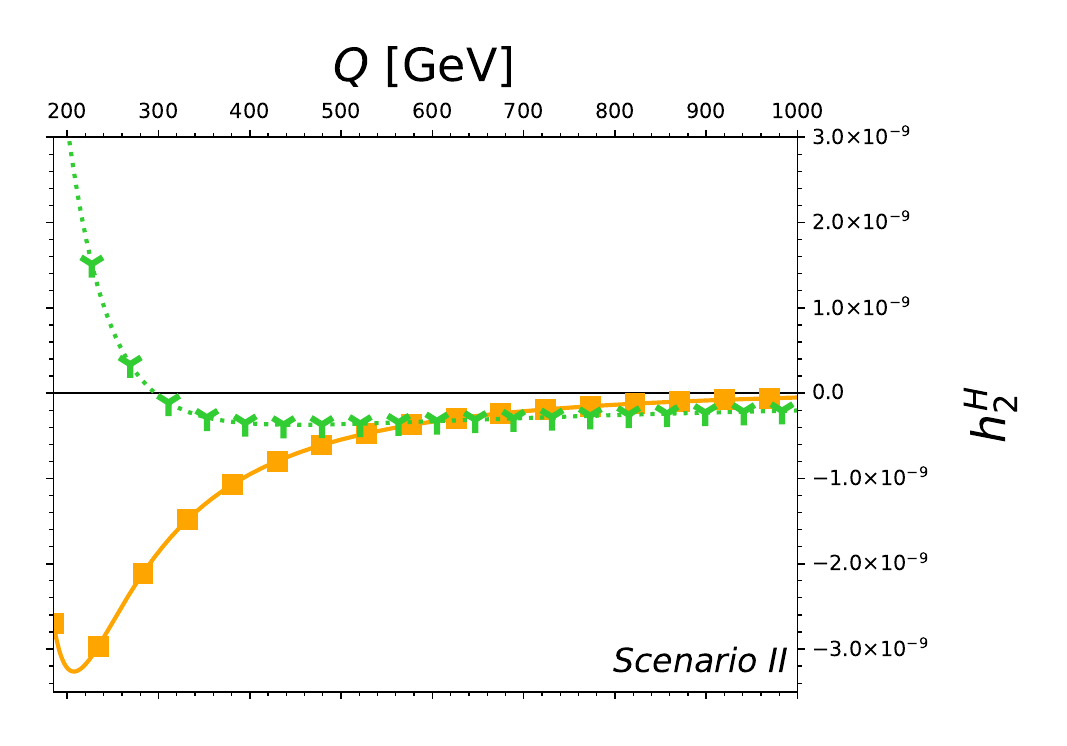}}\\\vspace{-.99cm}
\subfigure{\includegraphics[width=9cm]{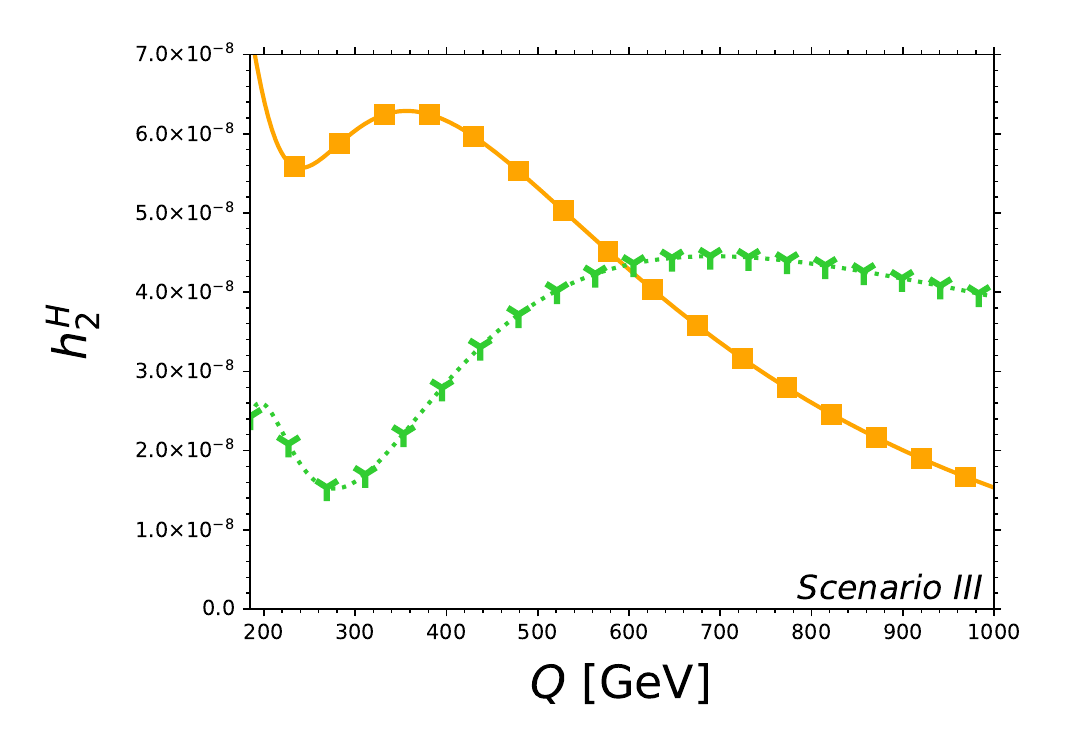}}\hspace{-.58cm}
\caption{FCNC contributions of the type II to the $h_2^H$ form factor as a function of $Q$. For the values of the different couplings, we consider the scenarios $I$-$III$, whereas the contribution in scenario $IV$ is tiny and not shown. We only consider the $Z\overline{t}c$ and $H\overline{t}c$ couplings.} \label{H*ZZplot2}
\end{center}
\end{figure}

In Fig. \ref{H*ZZplot2}, we show the behavior of $h_2^H(II)$ as a function of $Q$ for the scenarios $I-III$. At low values of $Q$, the dominant contributions correspond to the real part, while the absorptive part is of a similar order of magnitude. This behavior differs from that observed in contributions of type I, where the imaginary part becomes relevant only for energies above the threshold $Q=2m_t$. In the case of type II diagrams in Fig. \ref{diag2}, the Higgs boson is always coupled to a $\overline{t}c$ pair, which can be on-shell behind the threshold energy. Consequently, at low energies, the amplitudes of all the diagrams of type II develop an absorptive part, resulting in the real and imaginary parts being comparable in magnitude. As the energy increases, the imaginary part emerges as the dominant contribution. Notably, the most significant results are obtained in scenarios $I$ and $III$, of order $10^{-8}$. These values are three orders of magnitude larger than in the SM \cite{Hernandez-Juarez:2023dor} and two orders of magnitude less than the contributions from type I diagrams. Additionally, we observe distinct patterns in scenarios $II$ and $III$, indicating the significant impact of negative couplings on the behavior of $h_2^H(II)$. We do not present scenario IV, which involves only top quarks running in the loop, as $h_2^H$ is of order $10^{-20}$. 


\begin{figure}[H]
\begin{center}
\subfigure{\includegraphics[width=9cm]{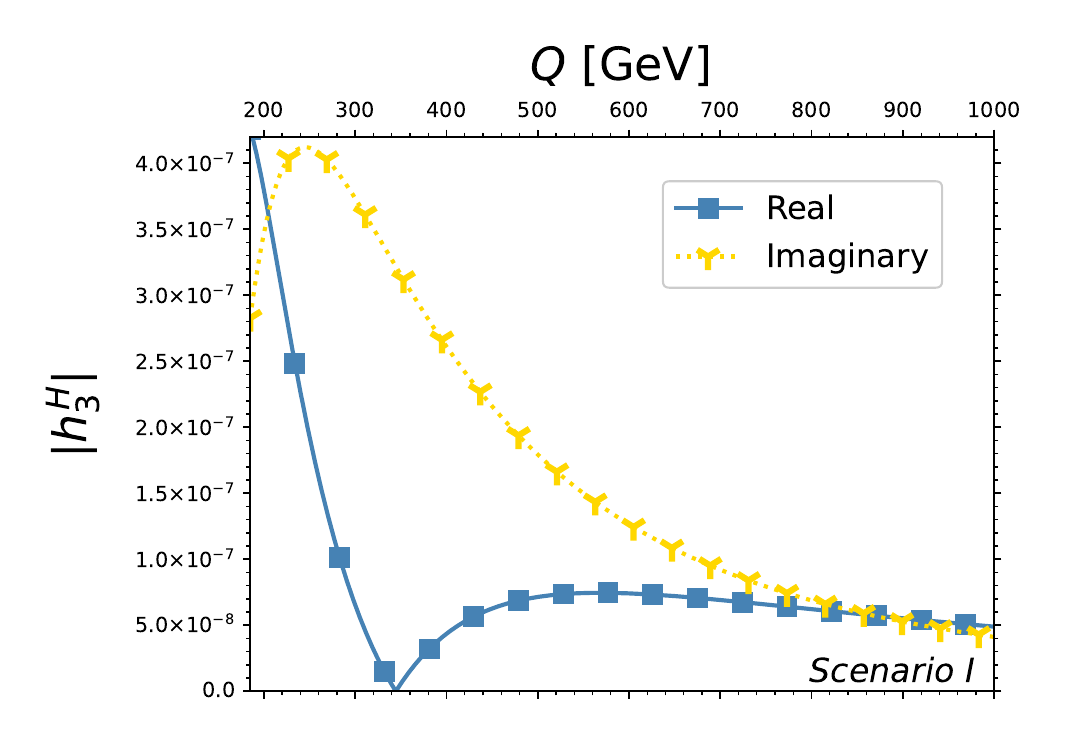}}\hspace{-.58cm}
\subfigure{\includegraphics[width=9cm]{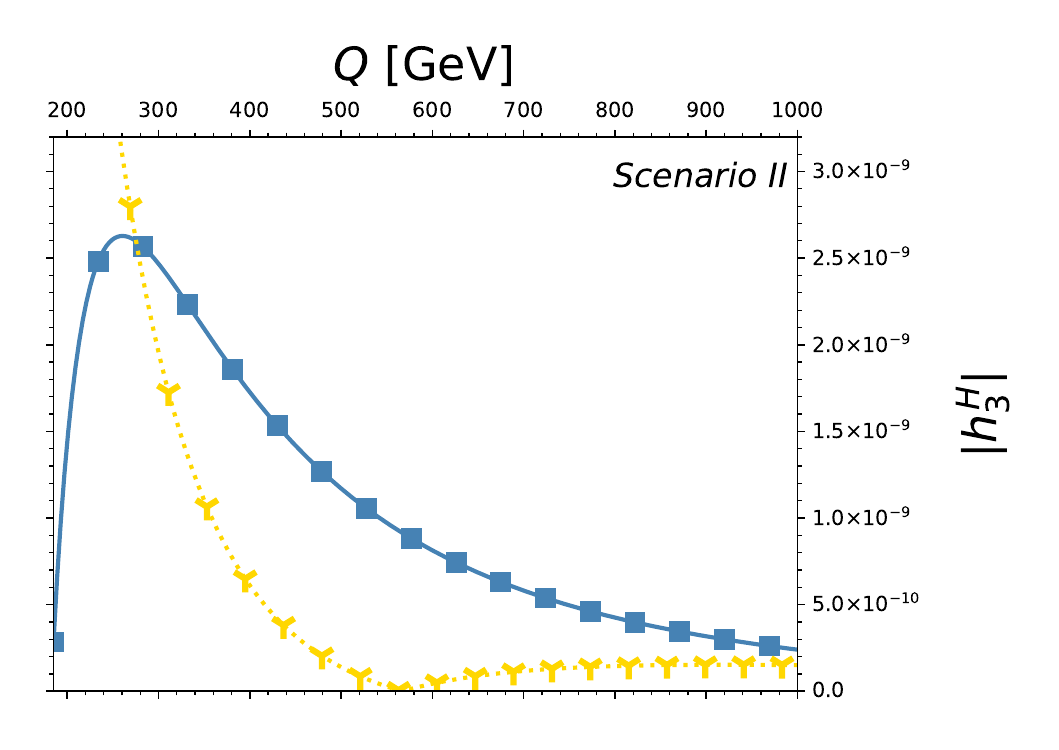}}\\\vspace{-.99cm}
\subfigure{\includegraphics[width=9cm]{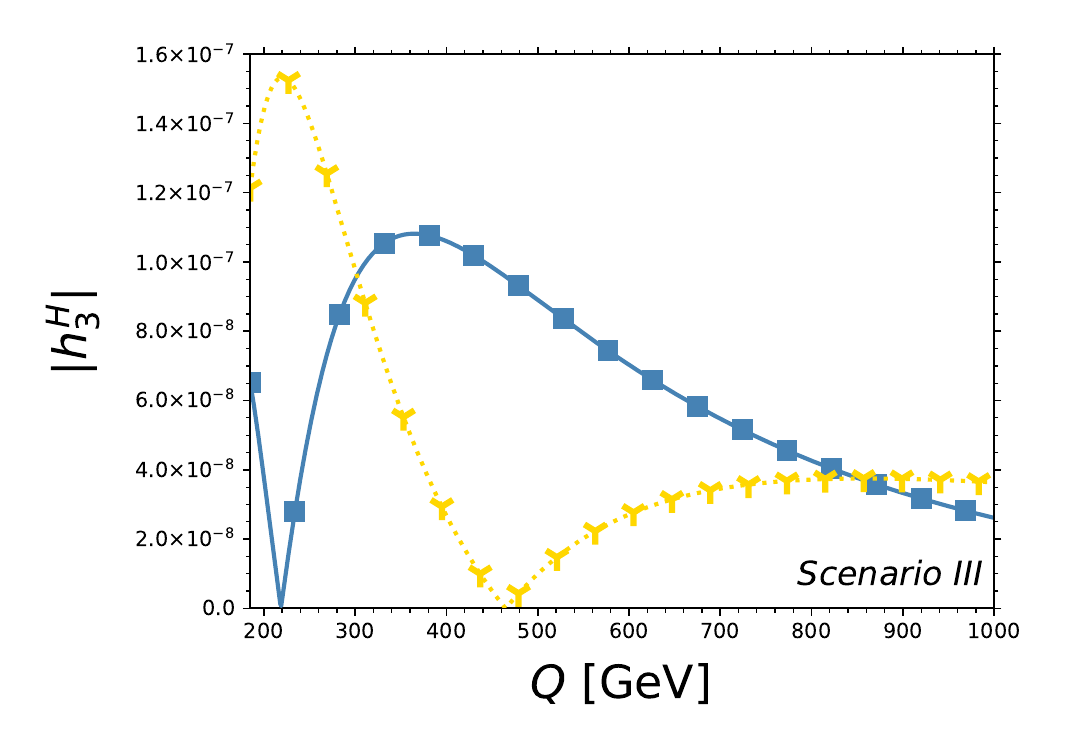}}\hspace{-.58cm}
\subfigure{\includegraphics[width=9cm]{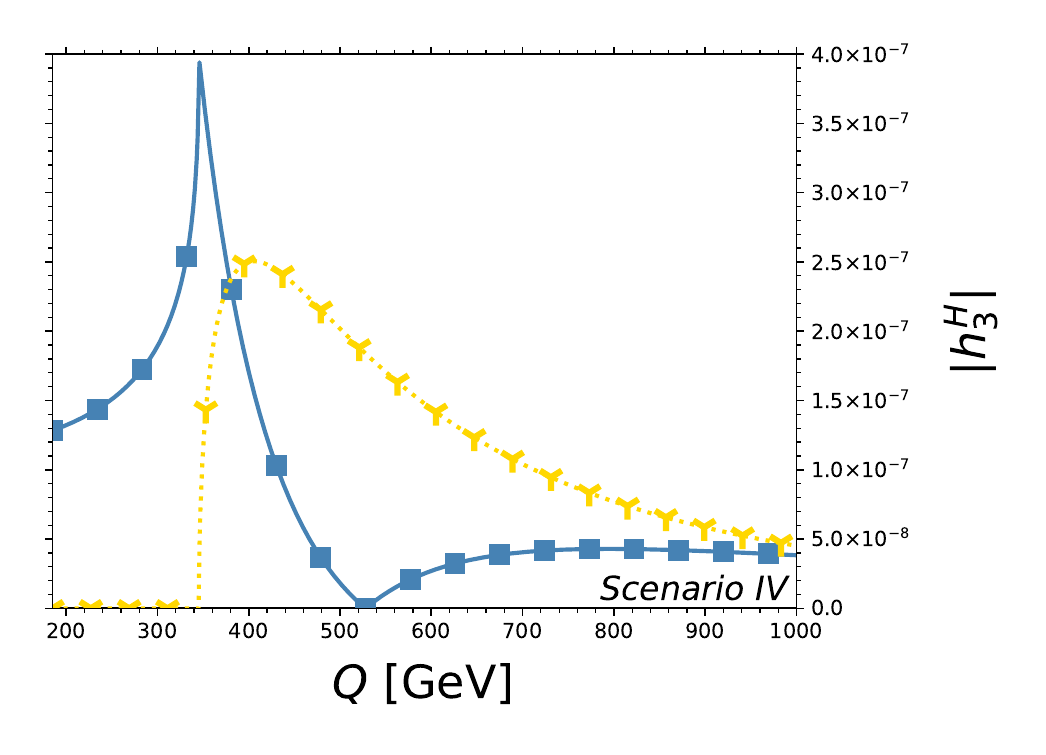}}
\caption{FCNC contributions of the type II to the $h_3^H$ form factor as a function of $Q$. For the values of the different couplings, we consider the scenarios $I$-$IV$ discussed in this section. We only consider the $Z\overline{t}c$ and $H\overline{t}c$ couplings.} \label{H*ZZplot3}
\end{center}
\end{figure}

For the $CP$-violating form factor, we present in Fig. \ref{H*ZZplot3} the absolute value of the real and absorptive part of $h_3^H(II)$ as a function of $Q$.  We observe that both contributions alternate in dominance across different energy regions. Furthermore, they decrease in magnitude and converge to similar values at high energies. In scenario $IV$, the absorptive part is zero for $Q<2m_t$, as only top quarks are included in the loop. We find the most significant values in scenarios $I$ and $IV$, where the real and imaginary parts can reach magnitudes of order $10^{-7}$. These values are four and one orders of magnitude larger than the prediction in the SM \cite{Soni:1993jc} and those obtained from type $I$ diagrams, respectively. Hence, relevant contributions to the left-right asymmetries from type II diagrams are feasible. Moreover, distinct patterns are evident in the four scenarios, indicating that negative couplings impact the behavior of $h_3^H(II)$. In contrast with the $CP$-conserving form factor $h^H_2(II)$, the contributions from the pseudoscalar coupling are significant for flavor-conserving scenarios as they are not negligible in scenario $IV$.

\subsubsection{$Z^\ast ZH$}

\begin{figure}[H]
\begin{center}
\subfigure{\includegraphics[width=9cm]{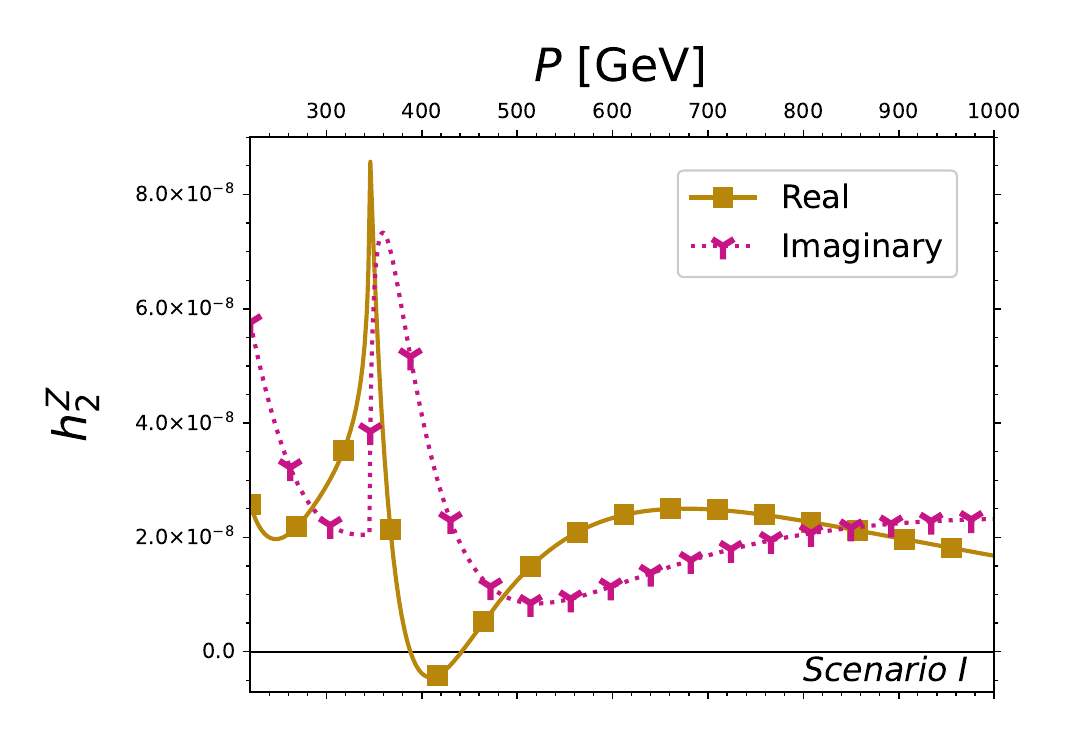}}\hspace{-.58cm}
\subfigure{\includegraphics[width=9cm]{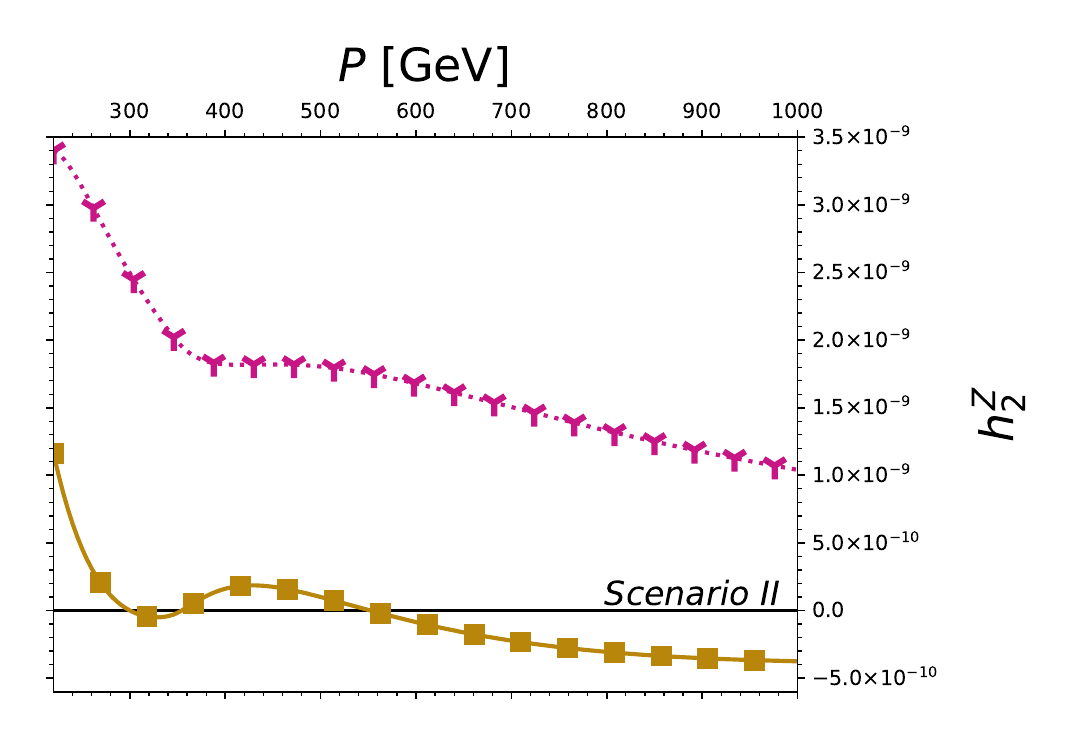}}\\\vspace{-.99cm}
\subfigure{\includegraphics[width=9cm]{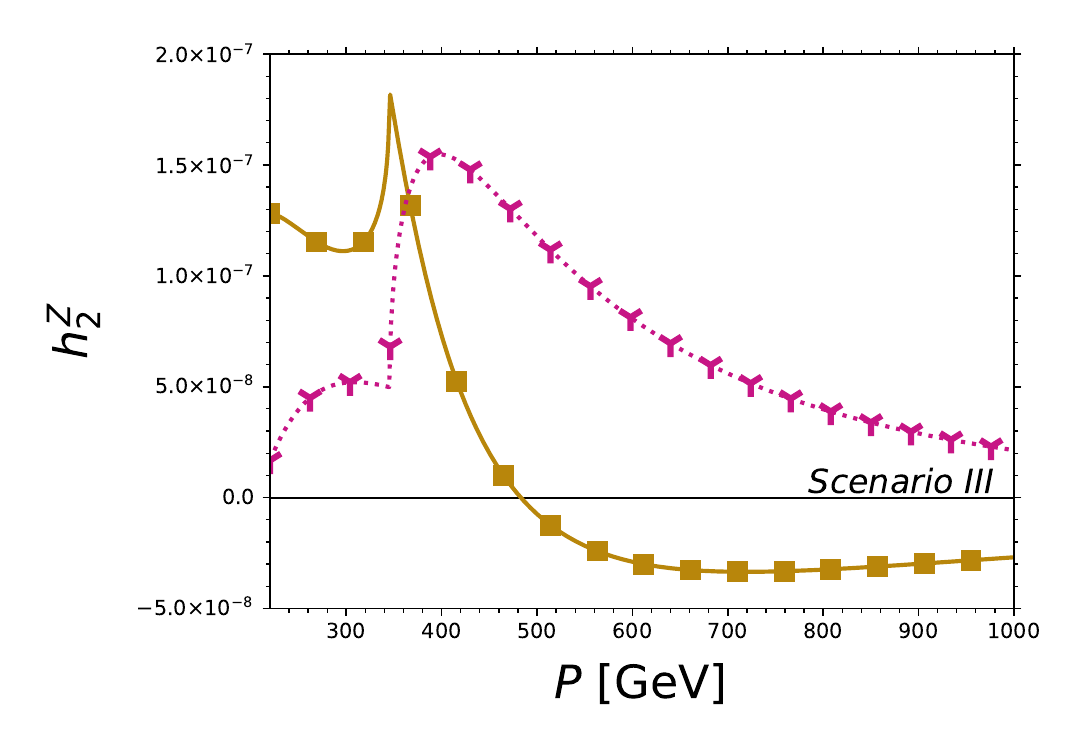}}\hspace{-.58cm}
\caption{FCNC contributions of the type II to the $h_2^Z$ form factor as a function of $P$. For the values of the different couplings, we consider the scenarios $I$-$III$, whereas the contribution in scenario $IV$ is tiny and not shown. We only consider the $Z\overline{t}c$ and $H\overline{t}c$ couplings.} \label{Z*ZHplot2}
\end{center}
\end{figure}

For the case of an off-shell $Z$ boson, we present the behavior of $h_2^Z(II)$ as a function of $P$ in Fig. \ref{Z*ZHplot2}. We note that the magnitudes of the real and absorptive parts are comparable, with the imaginary part dominating at high energies. The largest values occur in scenario $III$, reaching orders of $10^{-7}$. In scenario $IV$, the contributions are of order $10^{-20}$ and are not shown. Similar to the $H^\ast ZZ$ case, we find that negative couplings significantly influence the behavior of $h_2^Z(II)$. Our numerical results for $h_2^Z(II)$ are analogous to those with an off-shell Higgs boson. However, they are one order of magnitude smaller than those from contributions of type $I$ and are negligible compared to the SM result. 

\begin{figure}[H]
\begin{center}
\subfigure{\includegraphics[width=9cm]{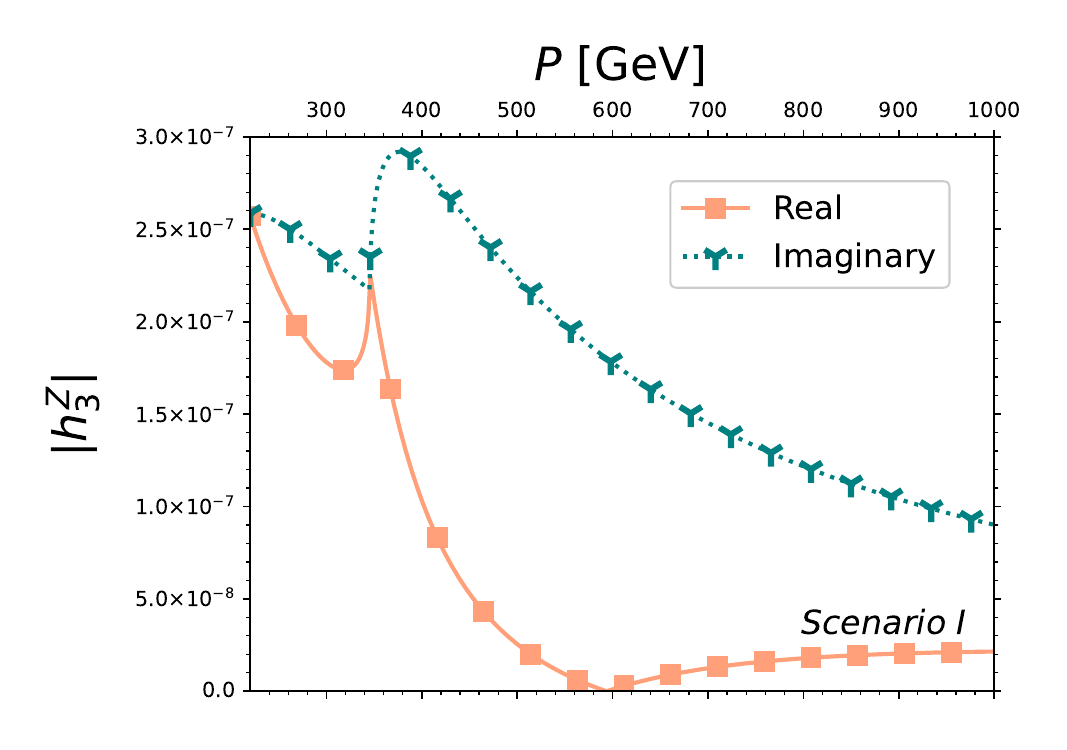}}\hspace{-.58cm}
\subfigure{\includegraphics[width=9cm]{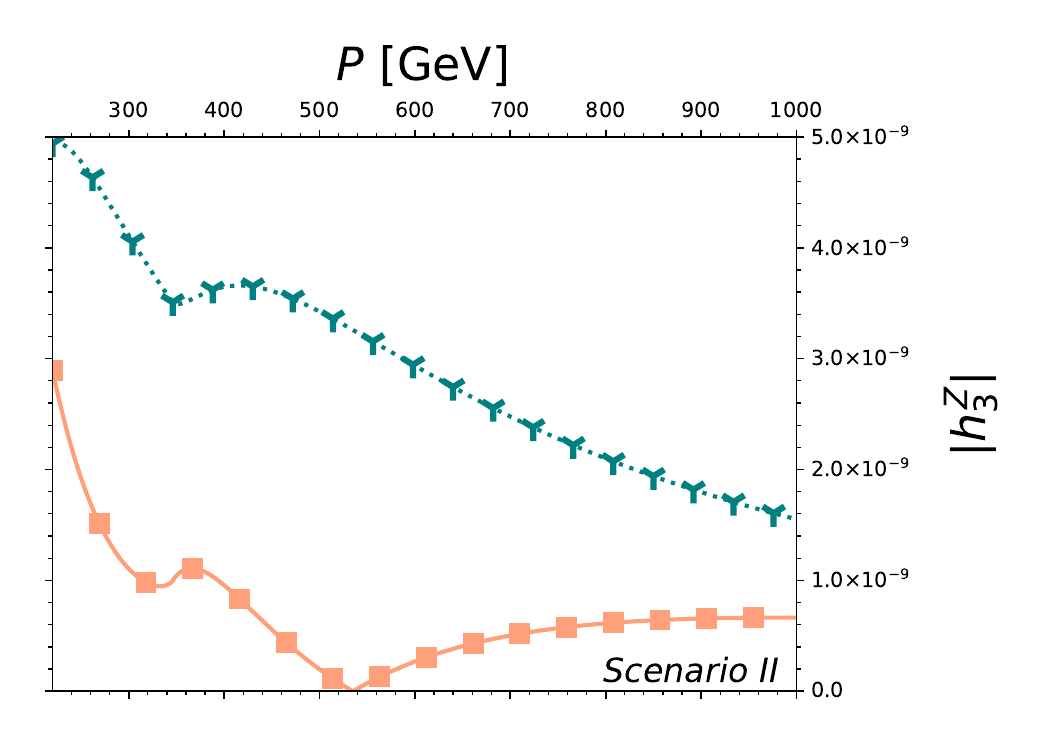}}\\\vspace{-.99cm}
\subfigure{\includegraphics[width=9cm]{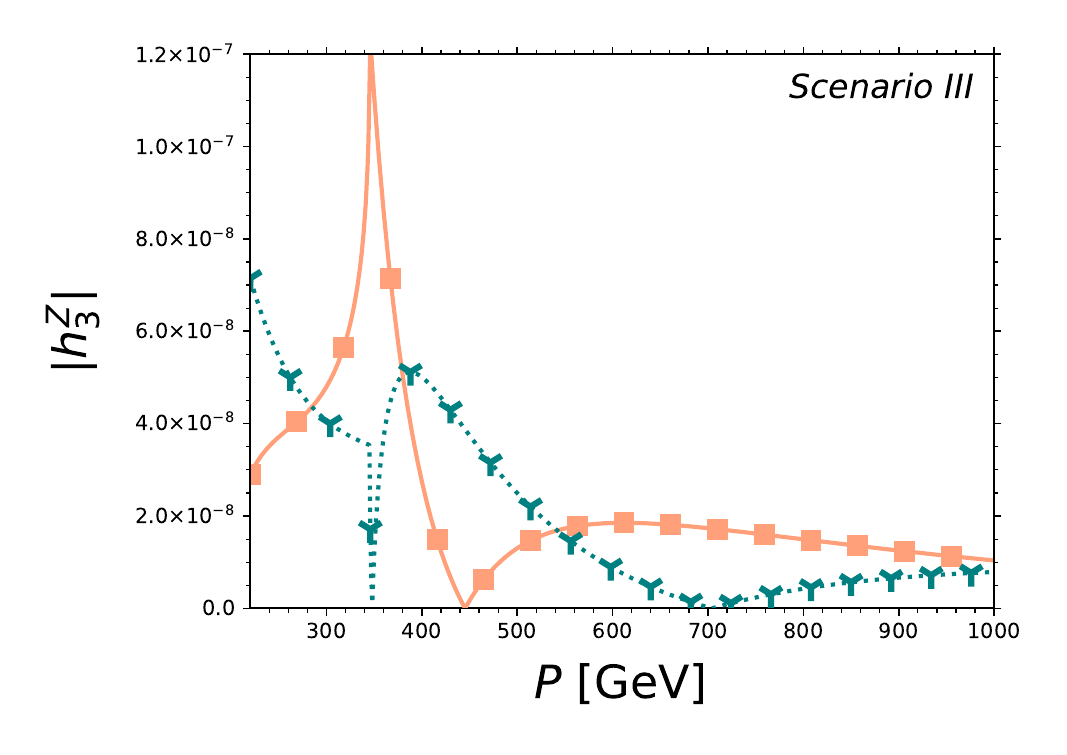}}\hspace{-.58cm}
\subfigure{\includegraphics[width=9cm]{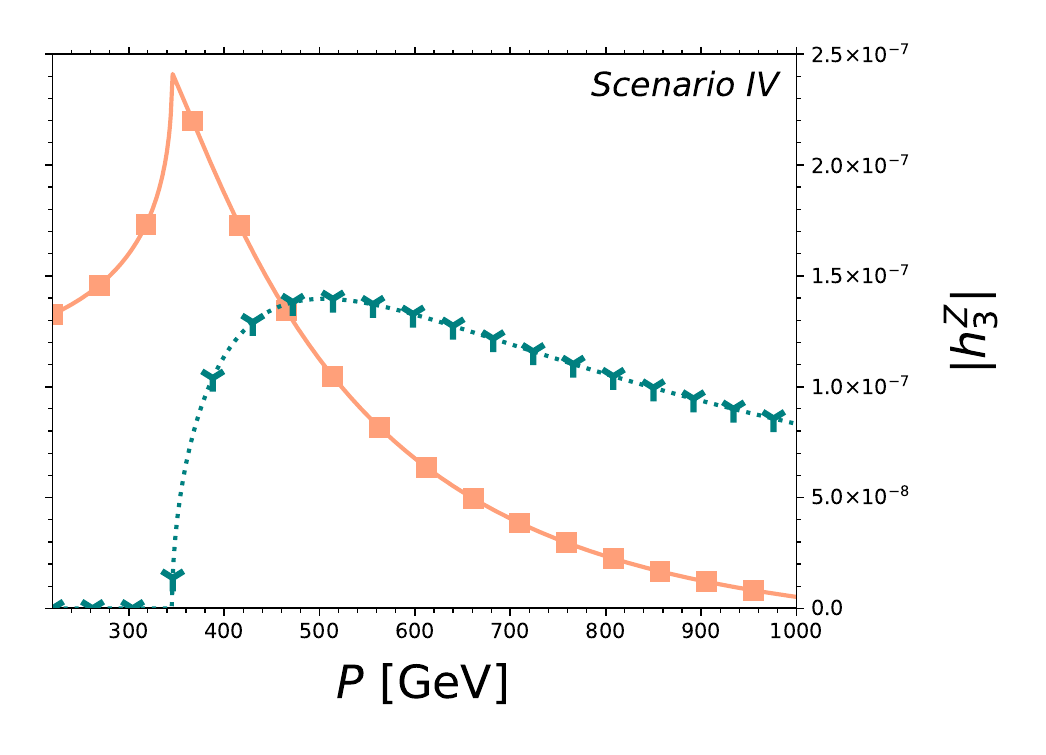}}
\caption{FCNC contributions of the type II to the $h_3^Z$ form factor as a function of $P$. For the values of the different couplings, we consider the scenarios $I$-$IV$ discussed in this section. We only consider the $Z\overline{t}c$ and $H\overline{t}c$ couplings.} \label{Z*ZHplot3}
\end{center}
\end{figure}

In Fig. \ref{Z*ZHplot3}, we show the absolute value of the real and imaginary parts of the $CP$-violating form factor $h_3^Z(II)$. In scenarios $I-II$, the imaginary part dominates, while in scenario $III$, the real part is the most significant contribution. In scenario $IV$, the absorptive part is zero for energies below $2m_t$, but it becomes the main contribution at higher energies. The real and absorptive parts of $h_3^Z(II)$ can reach values of order $10^{-7}$, four and one orders of magnitude larger than the SM predictions and those derived from type I diagrams, respectively. As observed in the $H^\ast ZZ$ vertex, the pseudoscalar coupling plays a relevant role in flavor-violating and flavor-conserving contributions to the $CP$-violating form factor $h_3^V$. An abrupt change at the threshold energy of $2m_t$ is noteworthy for both the real and imaginary parts, which differs from the behavior observed in Fig. \ref{H*ZZplot3}.


In summary, the imaginary and real parts of $h_2^V$ and $h_3^V$ ($V=H$, $Z$) are comparable for both contributions from diagrams type I and II. This behavior has also been observed in other off-shell couplings, such as trilinear neutral gauge bosons couplings \cite{Gounaris:2000tb, Hernandez-Juarez:2021mhi} and the $g\overline{q}q$ vertex \cite{Hernandez-Juarez:2020gxp, Hernandez-Juarez:2020drn}. The numerical values obtained for the $CP$-conserving form factor are negligible compared to those predicted in the SM and the current limits. In contrast, the contributions to $h_3^V$ can be up to five orders larger than in the SM. This form factor and the SM contributions to $h_1^V$ can lead to non-zero left-right asymmetries, as discussed in Sec. \ref{LRAsym}.

\subsection{Contributions to the $\mathcal{A}^V_{LR}$ ($V=H$, $Z$) asymmetry}


We now perform a numerical evaluation of the left-right asymmetries $\mathcal{A}_{LR}^V$ ($V=H$, $Z$) discussed in Sec. \ref{LRAsym}. 

 We note from Eqs. \eqref{AsymmetryH} and \eqref{AsymmetryZ} that only the form factors $h_1^V$ and $h_3^V$ are necessary for analyzing the behavior of the $\mathcal{A}_{LR}^V$ ($V=H$, $Z$) asymmetries. For the $CP$-conserving form factor in Eq. \eqref{H11}, we will only consider the tree-level and the SM one-loop contribution to $\hat{b}_Z$, as reported in Ref. \cite{Hernandez-Juarez:2023dor}. Regarding the $CP$-violating form factor $h_3^V$ ($V=H$, $Z$), we will utilize the expressions derived from FCNC contributions of both type I and II, discussed in Sec. \ref{secFCNCcon}.

\subsubsection{Type I}

\begin{figure}[H]
\begin{center}
\subfigure{\includegraphics[width=9cm]{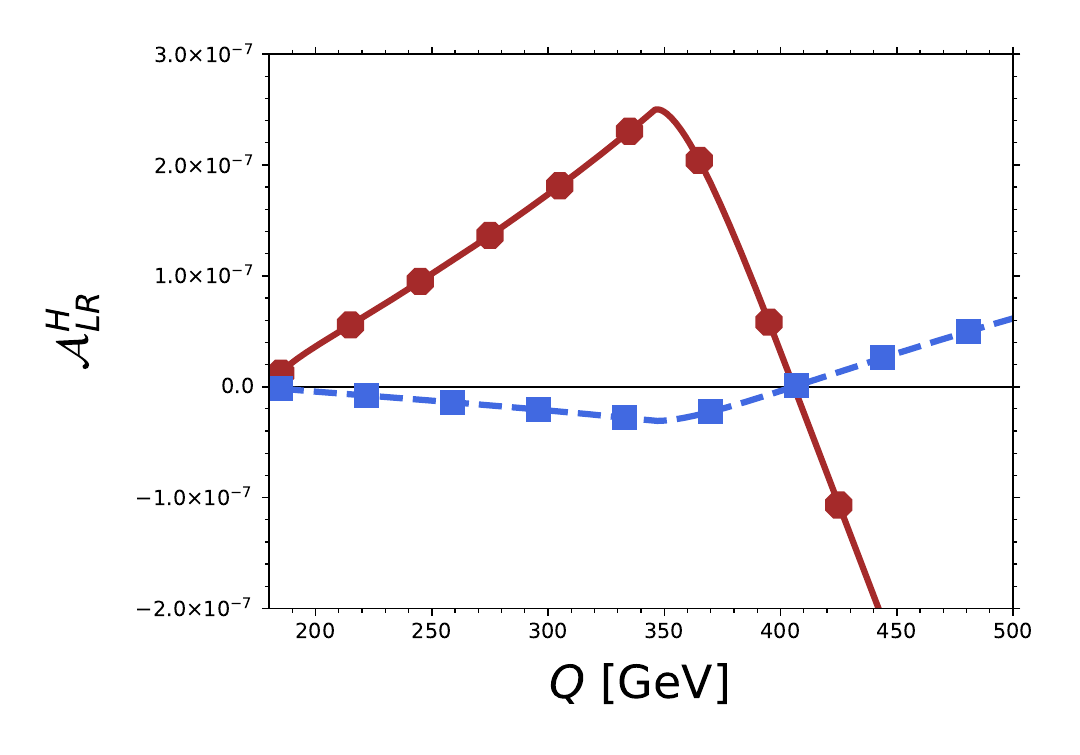}}\hspace{-.55cm}
\subfigure{\includegraphics[width=9cm]{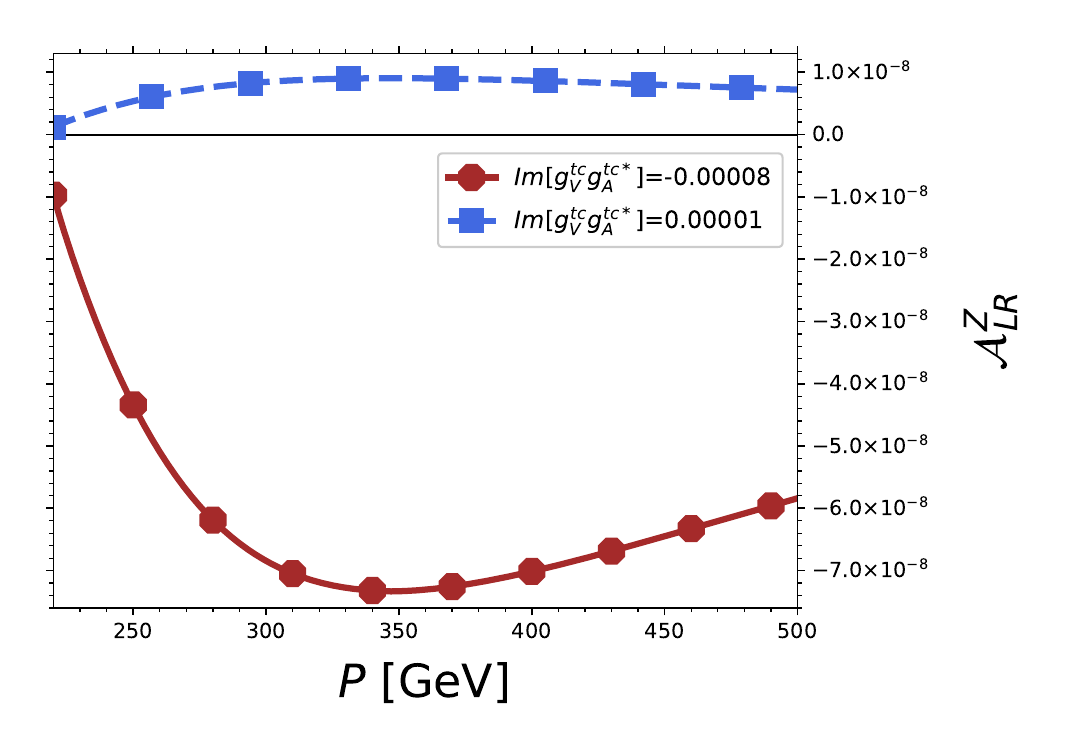}}\hspace{-.01cm}
\caption{The asymmetries $\mathcal{A}_{LR}^{H}$ and $\mathcal{A}_{LR}^{Z}$ as a function of $Q$ and $P$, respectively. For the $CP$-conserving form factor $h_2^V$ ($V=H$, $Z$), we have considered up the one-loop level contributions from the SM, whereas for $h_3^V$ we used the results obtained for FCNC contributions of type I. 
 } \label{as1}
\end{center}
\end{figure}

In Fig. \ref{as1}, we present the $\mathcal{A}_{LR}^{V}$ ($V=H$, $Z$) asymmetries as a function of $Q$ and $P$, considering the FCNC contributions of type I. Using the upper bounds stated in Eq. \eqref{bound1}, we find that ${\rm Im}\big[g^{tc}_V g^{tc}_A\big]$ can reach magnitudes of order $10^{-5}$. Accordingly, we utilize values of $-8\times10^{-5}$ and $10^{-5}$ for $ {\rm Im}\big[g^{tc}_V g^{tc}_A\big]$ in Eq. \eqref{h3type1}. The first value leads to the largest contributions for both asymmetries, of order $10^{-7}$ for the case involving an off-shell Higgs and $10^{-8}$ for the off-shell $Z$ boson. In contrast, the latter value of $ {\rm Im}\big[g^{tc}_V g^{tc}_A\big]$ results in asymmetries that are one order of magnitude smaller. Additionally, we observe distinct patterns between $\mathcal{A}_{LR}^H$ and $\mathcal{A}_{LR}^Z$. Both asymmetries exhibit an inflection point at $Q=2m_t$, where the one-loop SM contributions develop a notable imaginary part.

The SM prediction for the $CP$-violating form factor $h_3^H$ is aproximately of order $10^{-11}$ \cite{Soni:1993jc}. Based in this estimate, the $\mathcal{A}_{LR}^H$ asymmetry has been calculated to be of order $10^{-8}-10^{-9}$ \cite{Hernandez-Juarez:2023dor}. If we assume that the form factor $h_3^Z$ is similar in magnitude to that of an off-shell $H$ boson, we find that $\mathcal{A}_{LR}^Z$ can fall within the range $10^{-11}-10^{-12}$ in the SM. Thus, our results in Fig. \ref{as1} are one to three orders of magnitude larger than those predicted by the SM.

\subsubsection{Type II}

\begin{figure}[H]
\begin{center}
\subfigure{\includegraphics[width=9cm]{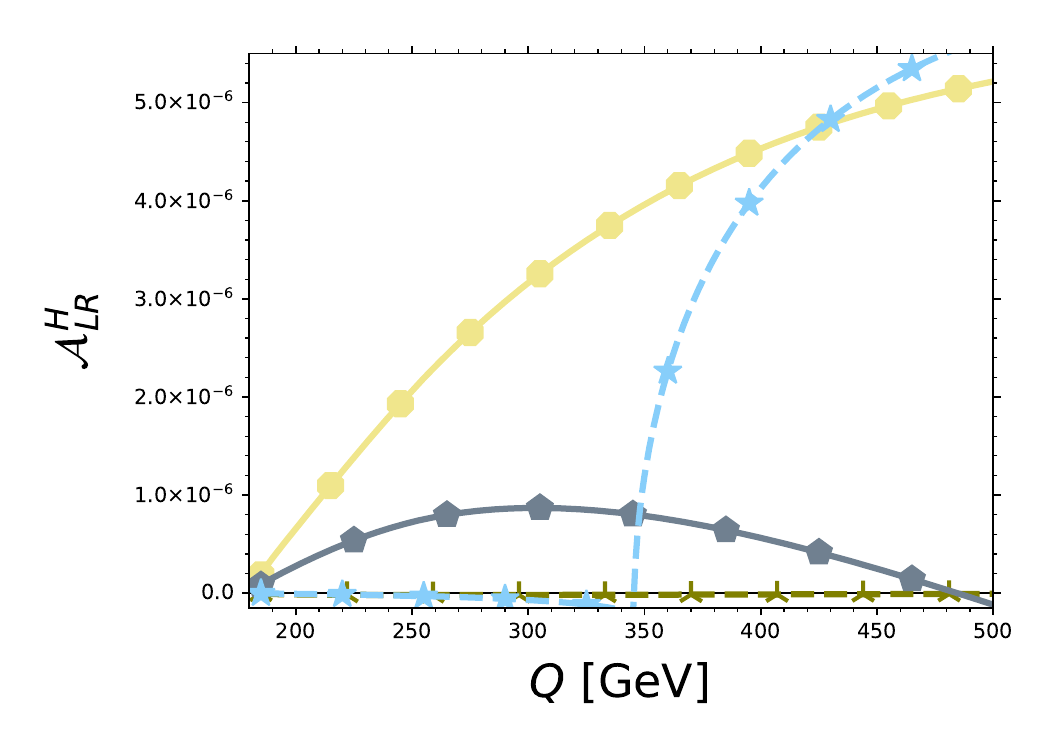}}\hspace{-.55cm}
\subfigure{\includegraphics[width=9cm]{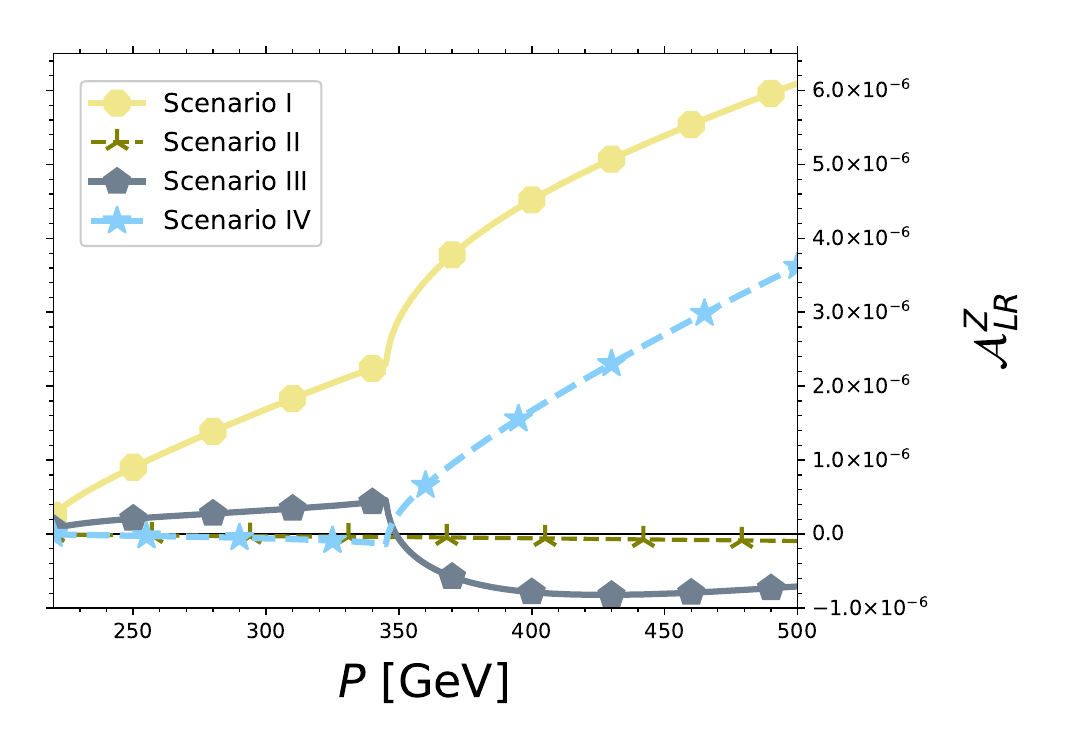}}\hspace{-.01cm}
\caption{The asymmetries $\mathcal{A}_{LR}^{H}$ and $\mathcal{A}_{LR}^{Z}$ as a function of $Q$ and $P$, respectively. For the $CP$-conserving form factor $h_2^V$ ($V=H$, $Z$), we have considered up the one-loop level contributions from the SM, whereas for $h_3^V$ we used the results obtained for FCNC contributions of type II.
} \label{as2}
\end{center}
\end{figure}

Similar to the previous case, we draw in Fig. \ref{as2} the $\mathcal{A}_{LR}^V$ ($V=H$, $Z$) asymmetries for the contributions of type II. We have examined the four scenarios introduced in Sec. \ref{subtypeIInum}. In both asymmetries, the scenario $II$ is negligible, while the remaining scenarios yield values ranging from $10^{-6}$ to $10^{-7}$. The most significant results are obtained in scenarios $I$ and $IV$, where $h_3^V(II)$ also reaches its highest values. Comparing both asymmetries, we observe that in scenarios $I$ and $III$, the $\mathcal{A}_{LR}^H$ asymmetry does not exhibit an inflection point at $Q=2m_t$. This phenomenon can be explained by the absence of any significant change in the real and imaginary parts of $h_3^H(II)$ at the threshold energy of $2m_t$, as illustrated in Fig. \ref{H*ZZplot3}. 

In scenario $IV$, the asymmetries become relevant at $Q=2m_t$, which aligns with the energy level where the $h_1^V$ and $h_3^V$ ($V=H$, $Z$) form factors also achieve considerable values. This finding indicates that the absorptive parts of the form factors notably influence the behavior of the asymmetries and should be considered in future analyses. We observe similar patterns between the two asymmetries, contrasting with the results obtained in Fig \ref{as1}.


 The $\mathcal{A}_{LR}^H$ and $\mathcal{A}_{LR}^Z$ asymmetries in Fig. \ref{as2} are one and two orders of magnitude larger than the corresponding from FCNC contributions of type I, respectively. Compared with the SM predictions, our results are greater by two to five orders of magnitude.

Finally, the feasibility of measuring the left-right asymmetries discussed in this note may be questioned. Recently, two methods have been developed to study polarized $ZZ$ final states at the LHC. The first method is the matrix-element reweighting method \cite{Maina:2021xpe}, which utilizes unpolarized events to obtain polarized observables through a fully differential reweighting technique. The second method can be easily integrated into existing public simulation tools and treats the polarizations ($\lambda$) of the $Z$ bosons as the propagators of new $V_\lambda$ bosons \cite{Javurkova:2024bwa}. In both methods, the definition of the polarized intermediate states is based on the completeness relationship:
\begin{equation}
\label{propa}
g_{\mu\nu}-\frac{q_\mu q_\nu}{M_V^2}=\sum_{\lambda}\epsilon_\mu(q,\lambda)\epsilon^\ast_\nu(q,\lambda), \quad \lambda=L, R, 0,
\end{equation}
which is introduced in the four-lepton decay amplitude by substituting the $Z$ bosons propagators. In the case of an off-shell $Z$ boson, this relationship can be extended by incorporating an auxiliary polarization $\epsilon_A^\mu$, which becomes zero when the boson is resonant \cite{Maina:2020rgd}. To derive the polarized observables from the unpolarized amplitude in the $H^\ast\to ZZ\to 4\ell$ process, one must consider only the amplitudes with on-shell $Z$. However, this selection breaks gauge invariance, which can be restored by employing the Narrow-Width- or Double-Pole Approximation \cite{Hoppe:2023uux}.

 The decay channel $H\to ZZ^\ast\to 4\ell$ has been studied at the LHC \cite{CMS:2019ekd, ATLAS:2020rej}. The signal region (SR) for this process is defined within the range of 115 GeV$<m_{4\ell}<$ 130 GeV, where $m_{4\ell}$ represents the four-lepton invariant mass. For the lepton pairs, the invariant masses must satisfy the conditions: 50 GeV $<m_{12}<$ 106 GeV and 12 GeV $<m_{34}<$ 115 GeV. The $m_{12}$ requirement ensures that the lepton pair originates from an on-shell $Z$ boson. Furthermore, the leading three leptons with the highest transverse momentum must have $p_T$ values exceeding 20 GeV, 15 GeV, and 10 GeV, respectively. The four leptons also require an angular separation of $\Delta R>0.10$. For the process $H^\ast\to ZZ\to 4\ell$, the SR is defined above the Higgs mass at 180 GeV$<m_{4\ell}$ \cite{ATLAS:2023dnm}. The same transverse momentum criteria of the leading three leptons established in the $H\to ZZ^\ast$ case are maintained. However, the four-lepton invariant masses conditions must satisfy  50 GeV $<m_{12}<$ 106 GeV and 50 GeV $<m_{34}<$115 GeV for 190 GeV $<m_{4\ell}$, ensuring that both $Z$ bosons are on-shell. These SRs have been used to assess the expected sensitivity to new physics contributions in polarized observables at the LHC \cite{Maina:2021xpe, Javurkova:2024bwa}. 
 
\begin{figure}[H]
\begin{center}
\subfigure{\includegraphics[width=9cm]{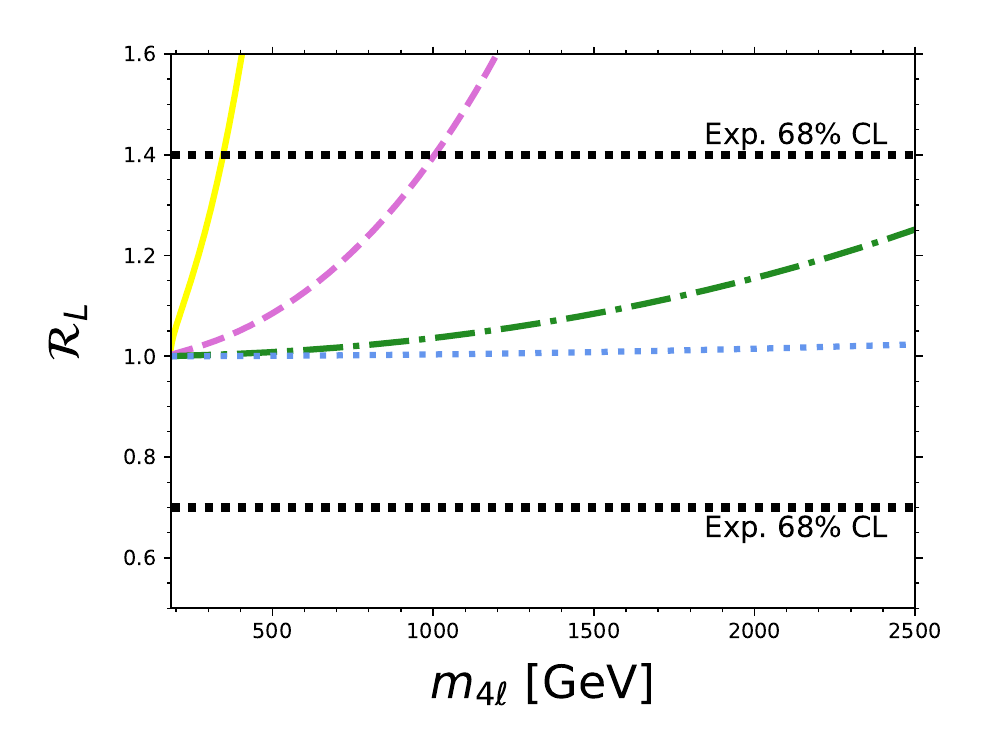}}\hspace{-.68cm}
\subfigure{\includegraphics[width=8.9cm]{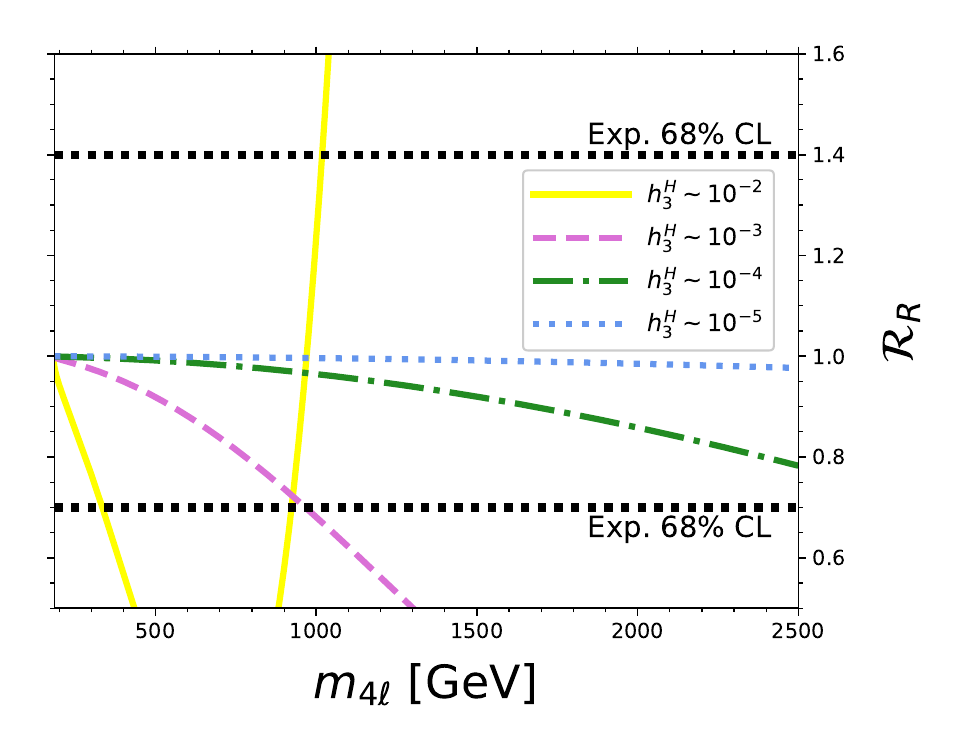}}\hspace{-.01cm}
\caption{Ratios between new physics and SM production of transversal polarizations as a function of the four-lepton invariant mass. We also plot the upper and lower expected limits at 68\% CL \cite{Javurkova:2024bwa}. 
} \label{sens}
\end{center}
\end{figure}

The observation of the left-right asymmetries necessitates precise measurements on polarized amplitudes. For Run 3 at the LHC, with an integrated luminosity of $\mathcal{L}=500$ fb$^{-1}$, the anticipated sensitivity to the ratios ($\mathcal{R}_{L, R}$) between new physics and SM production of transversal polarizations is projected to lie within 0.5$\lesssim \mathcal{R}_{L, R}\lesssim$1.7 (0.7$\lesssim \mathcal{R}_{L, R}\lesssim$1.5) at 95\% (68\%) CL, taking into account both systematic and statistical uncertainties associated with the $H^\ast\to ZZ$ process \cite{Javurkova:2024bwa}. In Fig. \ref{sens}, we show the behavior of the ratios $\mathcal{R}_{L, R}$ as a function of $m_{4\ell}$ for different values of $h_3^H$ within the range $10^{-2}-10^{-5}$. These values align with the current constraints on the $HZZ$ anomalous couplings \cite{CMS:2022ley, CMS:2022uox}. Notably, the $CP$-violating contributions in the left- and right-polarized amplitudes become more significant as the four-lepton invariant mass increases \cite{Hernandez-Juarez:2023dor}. We find that the effects corresponding to $h_3^H$ values of order $10^{-2}$ and $10^{-3}$ can be achieved at the 68\% CL for $m_{4\ell}$ less than 500 and 1000 GeV, respectively. For $h_3^H\sim10^{-4}$, invariant masses above 2500 GeV are required to distinguish the consequences of $CP$ violation. The projected sensitivity could be further improved using deep neural networks, similar to those implemented by the ATLAS or CMS  collaborations \cite{Javurkova:2024bwa}. Nevertheless, we estimate that to observe effects of $CP$-violating contributions of order $10^{-5}$ for low values of $m_{4\ell}$, the sensitivity on $\mathcal{R}_{L, R}$ would need to be increased by at least two orders of magnitude. 

As both theoretical and phenomenological methods for studying polarized processes continue to advance \cite{Ballestrero:2019qoy, Maina:2020rgd, Maina:2021xpe, Javurkova:2024bwa, Grossi:2024jae, Denner:2021csi, Pelliccioli:2023zpd, Dao:2023kwc, Dao:2024ffg, Carrivale:2025mjy}, we expect a significant enhancement in sensitivity to these observables soon. In this context, the asymmetries $\mathcal{A}_{LR}^V$ ($V=H$, $Z$) that arises from small $CP$-violating contributions may become measurable. The observation of a non-zero $\mathcal{A}^V_{LR}$ ($V=H$, $Z$) asymmetry would indicate the presence of new sources of $CP$ violation.

\section{conclusions}\label{concl}

In this work, we computed new contributions to the $H^\ast ZZ$ and $Z^\ast ZH$ vertices that arise from FCNC couplings mediated by the $Z$ and Higgs bosons. We identified two distinct categories of contributing Feynman diagrams. The first contribution, type I, results from considering only the FCNC couplings mediated by the $Z$ boson. The second contribution, type II, emerges when the FCNC couplings of the Higgs bosons are also included. From both scenarios, we found new contributions to the form factors $h_2^{V}$ and $h_3^V$ ($V=H$, $Z$), with the results expressed in terms of the Passarino-Veltman scalar functions. Notably, the $CP$-violating form factor $h_3^V$ can even be induced without flavor violation in a pseudoscalar coupling scenario. Using bounds on the FCNC couplings of the top quark, we determined that the contributions to $h_2^V$ can be of order $10^{-6}-10^{-7}$, which are significantly small compared with those in the SM. In contrast, the $CP$-violating form factor $h_3^V$ can reach values of $10^{-8}$ and $10^{-7}$ for type I and II contributions, respectively. These estimates are three and four orders of magnitude larger than the predictions of the SM. Additionally, we observed that the absorptive part is comparable to the real part and can dominate in certain energy regions. The contributions of FCNC mediated by the $Z$ and $H$ bosons to the $HZZ$ vertex have not been reported previously.

  Furthermore, we calculated the left-right asymmetry $\mathcal{A}_{LR}^Z$ for the $Z^\ast\to ZH$ process for the first time. Our results show that the $\mathcal{A}_{LR}^Z$ shares a similar form to the reported for the $H^\ast\to ZZ$ process, as both are in terms of the form factors $h_1^V$ and $h_3^V$. Using the SM one-loop level expressions for $h_1^V$ along with the contributions from FCNC couplings to $h_3^V$, we found that $\mathcal{A}_{LR}^H$  and $\mathcal{A}_{LR}^Z$  can reach significantly larger values than those derived from SM contributions alone. The most notable results arise from contributions of type II, where the asymmetries can be up to five orders of magnitude larger than in the SM. Additionally, we observe that the imaginary parts of the form factors considerably influence the behavior of the asymmetries. This finding suggests that the absorptive parts should not be ignored when studying physical observables. The detection of a non-zero left-right asymmetry $\mathcal{A}_{LR}^V$ ($V=H$, $Z$) would indicate the presence of $CP$ violation in the $HZZ$ vertex. However, the realization of such an observation requires enhanced sensitivity to polarized observables. We encourage studies focused on precise measurements of these observables.

\begin{acknowledgments}

This work  was supported by UNAM Posdoctoral Program (POSDOC), PAPIIT  project IN105825  "Estudios de f\'isica del sabor y violaci\'on de CP en modelos de nueva f\'isica", project CI2451 "B\'usqueda de nueva f\'isica en extensiones del Modelo Est\'andar". We also acknowledge support from  Sistema Nacional de Investigadores (Mexico). 
\end{acknowledgments}

\appendix
\section{Analytical Forms}
This appendix presents the analytical expressions for the functions appearing in the $h_2^V$ and $h_3^V$ ($V=H$, $Z$) form factors. Our results are in terms of the Passarino-Veltman scalar functions. We introduce the shorthand notation:
\begin{align}
    & B_{ij}(c^2)=B_0(c^2,m^2_i,m^2_j),  \\
    &  C_{ijk}(Q^2)=C_0(m_Z^2,m_Z^2,Q^2,m_i^2,m_j^2,m_k^2),\\
        &  C_{ijk}(P^2)=C_0(m_H^2,m_Z^2,P^2,m_i^2,m_j^2,m_k^2),
\end{align}
where $B_0$ and $C_0$ are the usual two- and three-point Passarino-Veltman scalar functions. The following symmetry relations will also be useful
\begin{align}
    & B_{ij}(c^2)=B_{ji}(c^2), \\
    &  C_{ijk}(c^2)=  C_{kji}(c^2).
\end{align}

\subsection{Diagrams type I}\label{Apptype1}

The functions $A^V_\mathcal{V}$ and $A^V_A$ ($V=H$, $Z$) for contributions of Type I are given as follows:

\begin{align}
\label{}
A^H_\mathcal{V}(Q^2,m_i^2,m_j^2) =&\frac{1}{Q^2\left(Q^2-4
   m_Z^2\right){}^2}\Bigg\{\Big[4 m_i^2 \left(m_i^2-m_j^2\right)
   \left(4 m_Z^2-Q^2\right)\Big] B_{ii}(0)+\Big[4 m_j^2 \left(m_j^2-m_i^2\right)
   \left(4 m_Z^2-Q^2\right)\Big]B_{jj}(0)\nonumber\\
   &+\Big[2 Q^4 \left(m_i-m_j\right){}^2-4
   Q^2 \big\{m_Z^2 \left(-4 m_i
   m_j+m_i^2+m_j^2\right)+\left(m_i
   ^2-m_j^2\right){}^2\big\}\nonumber\\
   &+8
   m_Z^4 \left(m_i^2+m_j^2\right)-8
   m_Z^2
   \left(m_i^2-m_j^2\right){}^2\Big] B_{ij}(m_Z^2)+ 2 m_i \Big[Q^4
   \left(m_j-m_i\right)\nonumber\\
   &+2 Q^2
   \big\{m_Z^2 \left(m_i-2
   m_j\right)+2 m_i
   \left(m_i-m_j\right)
   \left(m_i+m_j\right)\big\}-4
   m_i m_Z^2
   \left(m_i^2-m_j^2+m_Z^2\right)\Big] B_{ii}(Q^2)\nonumber\\
   &+2 m_j \Big[Q^4
   \left(m_i-m_j\right)+2 Q^2
   \big\{m_Z^2 \left(m_j-2
   m_i\right)+2 m_j
   \left(m_j-m_i\right)
   \left(m_j+m_i\right)\big\}\nonumber\\
   &-4
   m_j m_Z^2
   \left(m_j^2-m_i^2+m_Z^2\right)\Big]B_{jj}(Q^2)+m_i \Big[ 2 Q^4 \big\{-m_Z^2
   \left(m_i+3 m_j\right)-m_i^2
   m_j-3 m_i
   m_j^2\nonumber\\
   &+m_i^3+m_j^3\big\}+8 Q^2
   \big\{m_Z^2 \left(m_i^2 m_j+2
   m_i m_j^2+m_i^3-m_j^3\right)-m_i
   \left(m_i^2-m_j^2\right){}^2+m_j
   m_Z^4\big\}\nonumber\\
   &+8 m_i m_Z^2
   \left(m_i-m_j-m_Z\right)
   \left(m_i+m_j-m_Z\right)
   \left(m_i-m_j+m_Z\right)
   \left(m_i+m_j+m_Z\right)+Q^6
   m_j\Big]C_{iji}(Q^2)\nonumber\\
   &+m_j \Big[ 2 Q^4 \big\{-m_Z^2
   \left(m_j+3 m_i\right)-m_j^2
   m_i-3 m_j
   m_i^2+m_j^3+m_i^3\big\}+8 Q^2
   \big\{m_Z^2 \big(m_j^2 m_i+2
   m_j m_i^2\nonumber\\
   &+m_j^3-m_i^3\big)-m_j
   \left(m_j^2-m_i^2\right){}^2+m_i
   m_Z^4\big\}+8 m_j m_Z^2
   \left(m_j-m_i-m_Z\right)
   \left(m_j+m_i-m_Z\right)\nonumber\\
   &\times
   \left(m_j-m_i+m_Z\right)
   \left(m_j+m_i+m_Z\right)+Q^6
   m_i\Big]C_{jij}(Q^2)\nonumber\\
   &+2 \Big[4 m_Z^2-Q^2\Big]
   \Big[m_i^2 \left(-4 m_j^2-2
   m_Z^2+Q^2\right)+2 m_i^4+m_j^2
   \left(2 m_j^2-2
   m_Z^2+Q^2\right)\Big]\Bigg\}
   \end{align}

\begin{align}
\label{}
A^H_A(Q^2,m_i^2,m_j^2) =&A^H_\mathcal{V}(Q^2,m_i^2,m_j^2) -\frac{1}{Q^2\left(Q^2-4
   m_Z^2\right){}^2}2 Q^2 m_i m_j \left(Q^2-4
   m_Z^2\right)\Big[ -4
   B_{ij}\left(m_Z^2\right)+2
   B_{ii}\left(Q^2\right)+2
   B_{jj}\left(Q^2\right)\nonumber\\
   &+\big\{Q^2-2
   \left(m_i^2-m_j^2+m_Z^2\right)\big\}
   C_{ij}\left(Q^2\right)+\big\{2
   m_i^2-2
   \left(m_j^2+m_Z^2\right)+Q^2\big\}
   C_{ji}\left(Q^2\right) \Big]
\end{align}

\begin{align}
\label{}
A^Z_\mathcal{V}(P^2,m_i^2,m_j^2) =&\frac{1}{\left((m_H-m_Z)^2-P^2\right)
   \left((m_H+m_Z)^2-P^2\right)
   }\Bigg\{ -4 m_i^2 \left(m_i-m_j\right)
   \left(m_i+m_j\right)B_{ii}(0)\nonumber\\
   &+m_i \Bigg[\frac{3 m_H m_i \left(m_Z
   \left(m_H+m_Z\right)+m_i^2-m_j^2
   \right)}{m_H+m_Z-P}+\frac{3
   m_H m_i \left(m_Z
   \left(m_Z-m_H\right)+m_i^2-m_j^2
   \right)}{m_H-m_Z+P}\nonumber\\
   &-\frac{3
   m_H m_i \left(m_Z
   \left(m_Z-m_H\right)+m_i^2-m_j^2
   \right)}{-m_H+m_Z+P}+\frac{3
   m_H m_i \left(m_Z
   \left(m_H+m_Z\right)+m_i^2-m_j^2
   \right)}{m_H+m_Z+P}\nonumber\\
   &-2
   \left(m_i \left(m_H^2-2
   m_j^2+m_Z^2\right)-m_H^2 m_j+2
   m_i^3\right)-2 P^2 m_i\Bigg]B_{ii}(m_H^2)\nonumber\\
   &+\Bigg[ \frac{3 m_Z \left(-m_H m_Z
   \left(m_i^2+m_j^2\right)+m_Z^2
   \left(m_i^2+m_j^2\right)+\left(m
   _i^2-m_j^2\right){}^2\right)}{-m
   _H+m_Z-P}\nonumber\\
   &+\frac{3 m_Z
   \left(m_i^2 \left(m_Z
   \left(m_H+m_Z\right)-2
   m_j^2\right)+m_j^2 \left(m_Z
   \left(m_H+m_Z\right)+m_j^2\right)+m_i^4\right)}{m_H+m_Z-P}\nonumber\\
   &+\frac{3 m_Z \left(-m_H m_Z
   \left(m_i^2+m_j^2\right)+m_Z^2
   \left(m_i^2+m_j^2\right)+\left(m
   _i^2-m_j^2\right){}^2\right)}{-m
   _H+m_Z+P}\nonumber\\
   &+\frac{3 m_Z
   \left(m_i^2 \left(m_Z
   \left(m_H+m_Z\right)-2
   m_j^2\right)+m_j^2 \left(m_Z
   \left(m_H+m_Z\right)+m_j^2\right)+m_i^4\right)}{m_H+m_Z+P}-2
   m_H^2 m_i m_j\nonumber\\
   &+m_H^2 m_i^2+m_H^2
   m_j^2-P^2
   \left(m_i-m_j\right){}^2-2 m_i
   m_j m_Z^2+4 m_i^2 m_j^2-9 m_i^2
   m_Z^2-2 m_i^4-9 m_j^2 m_Z^2\nonumber\\
   &-2
   m_j^4\Bigg]\frac{B_{ij}(m_Z^2)}{2} +\frac{1}{ \left((m_H-m_Z)^2+P^2\right)
   \left((m_H+m_Z)^2-P^2\right)
   }\Big[-2 m_i m_j
   \left((m_H-m_Z)^2-P^2\right)\nonumber\\
   &\times
   \left((m_H+m_Z)^2-P^2\right)
   \left(m_H^2-m_Z^2+P^2\right)+m_i
   ^2 \big(-5 P^4 \left(m_H^2+4
   m_j^2-m_Z^2\right)\nonumber\\
   &+P^2 \left(16
   m_j^2 \left(m_H^2+m_Z^2\right)+6
   m_H^2 m_Z^2+m_H^4-7
   m_Z^4\right)+\left(m_H^2-m_Z^2\right){}^2 \left(m_H^2+4
   m_j^2-m_Z^2\right)+3
   P^6\big)\nonumber\\
   &-2 m_i^4 \left(4 P^2
   \left(m_H^2+m_Z^2\right)+\left(m
   _H^2-m_Z^2\right){}^2-5
   P^4\right)+m_j^2 \big(5 P^4
   \left(-m_H^2+2
   m_j^2+m_Z^2\right)\nonumber\\
   &+P^2 \left(-8
   m_j^2 \left(m_H^2+m_Z^2\right)+6
   m_H^2 m_Z^2+m_H^4-7
   m_Z^4\right)+\left(m_H^2-m_Z^2\right){}^2 \left(m_H^2-2
   m_j^2-m_Z^2\right)\nonumber\\
   &+3
   P^6\big)\Big]\frac{B_{ij}(P^2)}{2}+ \frac{m_i }{\
   \left((m_H-m_Z)^2-P^2\right)
   \left((m_H+m_Z)^2-P^2\right)}\Big[2 m_i^3 \big(m_j^2
   \big(-4 m_H^2
   \left(m_Z^2+P^2\right)\nonumber\\
   &+8 m_H^4-4
   \left(m_Z^2-P^2\right){}^2\big)+2 m_H^4
   \left(m_Z^2+P^2\right)+m_H^2
   \left(6 P^2 m_Z^2-7 m_Z^4-7
   P^4\right)+m_H^6\nonumber\\
   &+4
   \left(m_Z^2-P^2\right){}^2
   \left(m_Z^2+P^2\right)\big)+m_
   i \big(4 m_j^4 \left(m_H^2
   \left(m_Z^2+P^2\right)-2
   m_H^4+\left(m_Z^2-P^2\right){}^2
   \right)\nonumber\\
   &+2 m_j^2 \left(4 m_H^4
   \left(m_Z^2+P^2\right)+m_H^2
   \left(-10 P^2
   m_Z^2+m_Z^4+P^4\right)-3 m_H^6-2
   \left(m_Z^2-P^2\right){}^2
   \left(m_Z^2+P^2\right)\right)\nonumber\\
   &+\left(-m_H^2+m_Z^2+P^2\right)
   \big(-P^4 \left(2
   m_H^2+m_Z^2\right)+P^2 \left(8
   m_H^2
   m_Z^2+m_H^4-m_Z^4\right)+\left(m
   _Z^3-m_H^2
   m_Z\right){}^2\nonumber\\
   &+P^6\big)\big)
   -2 m_H^2 m_i^2 m_j
   \left((m_H-m_Z)^2-P^2\right)
   \left((m_H+m_Z)^2-P^2\right)
+4 m_i^5
   \big(m_H^2
   \left(m_Z^2+P^2\right)\nonumber\\
   &-2
   m_H^4+\left(m_Z^2-P^2\right){}^2
   \big)+m_H^2 m_j
   \left((m_H-m_Z)^2-P^2\right)
   \left((m_H+m_Z)^2-P^2\right)
   \big(m_H^2+2
   m_j^2-m_Z^2\nonumber\\
   &-P^2\big)\Big]C_{iij}(P^2)+ \left(m_i^2 \left(-m_H^2+4
   m_j^2+m_Z^2+P^2\right)+m_j^2
   \left(-m_H^2-2
   m_j^2+m_Z^2+P^2\right)-2
   m_i^4\right)\nonumber\\
   &+(i\leftrightarrow j)\Bigg\}.
\end{align}

\begin{align}
\label{}
A^Z_A(p^2,m_i^2,m_j^2) =&A^Z_\mathcal{V}(P^2,m_i^2,m_j^2)-\frac{2 m_i m_j}{\left((m_H-m_Z)^2-P^2\right)
   \left((m_H+m_Z)^2-P^2\right)
   } \Big[-2 m_H^2\big\{
   B_{ij}\left(P^2\right)+
   B_{ij}\left(m_Z^2\right)\big\}\nonumber\\
   &+2 m_H^2
   \big\{B_{ii}\left(m_H^2\right)+
   B_{jj}\left(m_H^2\right)\big\}+2 P^2\Big\{
   B_{ij}\left(m_Z^2\right)-
   B_{ij}\left(P^2\right)\Big\}+2 m_Z^2\Big\{
B_{ij}\left(P^2\right)-
B_{ij}\left(m_Z^2\right)\Big\}\nonumber\\
&+m_H^2\big(m_H^2
-2 
   m_i^2
   +2 
   m_j^2-
   m_Z^2-P^2
   \big)C_{iij}\left(P^2\right)+m_H^2
   \left(m_H^2+2 m_i^2-2
   m_j^2-m_Z^2-P^2\right)
C_{jji}\left(P^2\right)\Big]
\end{align}

 The functions $\widetilde{\mathcal{F}}^V$ ($V=H$, $Z$) for contributions of Type I are expressed as follows:

\begin{align}
\label{}
\widetilde{\mathcal{F}}^H(Q^2,m_i,m_j)=&\frac{1}{\left(4
   m_Z^2-Q^2\right)}\Bigg\{2\Big[
   B_{jj}\left(Q^2\right)-
   B_{ii}\left(Q^2\right)\Big]+\Big[2
   \left(m_i^2-m_j^2+m_Z^2\right)-Q
   ^2\Big]
   C_{iji}\left(Q^2\right)\nonumber\\
   &+\Big[2
   \left(m_i^2-m_j^2-m_Z^2\right)+Q
   ^2\Big]
   C_{jij}\left(Q^2\right)\Bigg\},
\end{align}

\begin{align}
\label{}
\widetilde{\mathcal{F}}^Z(Q^2,m_i,m_j)=&-\frac{  m_H^2 }
   {   \left((m_H-m_Z)^2-P^2\right)
   \left((m_H+m_Z)^2-P^2\right)}
   \Bigg\{2\Big[
   \text{B}_{jj}\left(m_H^2\right)-
   \text{B}_{ii}\left(m_H^2\right)\Big]\nonumber\\
   &+\left(-
   m_H^2+2 m_i^2-2
   m_j^2+m_Z^2+P^2\right)
C_{iij}\left(P^2\right)+\left(m_H^2+2 m_i^2-2
   m_j^2-m_Z^2-P^2\right)
C_{jji}\left(P^2\right)\Bigg\}
\end{align}

\subsection{Diagrams type II}
\label{Apptype2}

 The functions $R^V_{1,2,3,4}$ ($V=H$, $Z$) for contributions of Type II are given as follows:

\begin{align}
\label{}
 R^H_1(Q^2,m_i,m_j)=&\frac{1}{Q^4\left(Q^2-4 m_Z^2\right){}^2} \Bigg\{m_Z^2\Big[6 m_i^2 m_j \left(Q^2-2 m_Z^2\right)-m_i \left(6 m_j^2 \left(Q^2-2
   m_Z^2\right)-10 Q^2 m_Z^2+Q^4\right)\nonumber\\
   &+6 m_i^3 \left(Q^2-2 m_Z^2\right)+3 m_j
   \left(Q^2-2 m_j^2\right) \left(Q^2-2 m_Z^2\right)\Big]B_{ii}(m_Z^2)\nonumber\\
   &+  Q^2  m_Z^2 \left(m_i+m_j\right) \left(2 m_Z^2+Q^2\right)\Big[B_{ij}(m_Z^2)-2 B_{ij}(Q^2)\Big]\nonumber\\
   &+ \Big[-6 m_i^4 m_j m_Z^2 \left(Q^2-2 m_Z^2\right)-2 m_i^3 \left(Q^6-m_Z^2 \left(6
   m_j^2+5 Q^2\right) \left(Q^2-2 m_Z^2\right)\right)\nonumber\\
   &-2 m_i^2 m_j \left(Q^2-2
   m_Z^2\right) \left(Q^4-m_Z^2 \left(6 m_j^2+Q^2\right)\right)+m_i \big(4 m_Z^4
   \left(Q^2 m_j^2+3 m_j^4+2 Q^4\right)\nonumber\\
   &+m_Z^2 \left(2 Q^4 m_j^2-6 Q^2 m_j^4-7
   Q^6\right)+4 Q^2 m_Z^6+Q^8\big)-6 m_i^5 m_Z^2 \left(Q^2-2 m_Z^2\right)\nonumber\\
   &+m_j m_Z^2
   \left(2 m_Z^2-Q^2\right) \left(-6 Q^2 m_j^2+6 m_j^4+2 Q^2 m_Z^2+Q^4\right)\Big]C_{iij}(Q^2)\nonumber\\
   &- Q^2  \left(m_i+m_j\right) \left(-6 Q^2 m_Z^2+8 m_Z^4+Q^4\right)\Bigg\}+ (i \leftrightarrow j),
\end{align}

\begin{align}
\label{}
 R^H_2(Q^2,m_i,m_j)= &\frac{1}{Q^4\left(Q^2-4 m_Z^2\right){}^2} \Bigg\{\left(Q^2-2 m_Z^2\right) \big(m_Z^2 \left(-m_i \left(6 m_j^2+5 Q^2\right)-6 m_i^2
   m_j+6 m_i^3-3 Q^2 m_j+6 m_j^3\right)\nonumber\\
   &+2 Q^4 m_i\big)B_{ii}(m_Z^2)+Q^2  \left(m_i-m_j\right) \left(-3 Q^2 m_Z^2+2 m_Z^4+Q^4\right)\Big[B_{ij}(m_Z^2) -2  B_{ij}(Q^2)\Big]\nonumber\\
   &- \left(Q^2-2 m_Z^2\right) \big(2 Q^2 m_Z^4 \left(m_i-m_j\right)+m_Z^2 \big(2 m_i
   m_j^2 \left(Q^2-6 m_i^2\right)+6 m_j^3 \left(2 m_i^2+Q^2\right)\nonumber\\
   &-m_j \left(-2 Q^2
   m_i^2+6 m_i^4+Q^4\right)+6 m_i m_j^4-3 Q^4 m_i-10 Q^2 m_i^3+6 m_i^5-6
   m_j^5\big)\nonumber\\
   &+Q^4 m_i \left(2 \left(m_i-m_j\right) \left(2
   m_i+m_j\right)+Q^2\right)\big)C_{iij}(Q^2)\nonumber\\
   &- Q^2  \left(m_i-m_j\right) \left(-6 Q^2 m_Z^2+8 m_Z^4+Q^4\right)\Bigg\}-(i \leftrightarrow j),
\end{align}

\begin{align}
\label{}
 R^H_3(Q^2,m_i,m_j)=&\frac{1}{Q^4\left(Q^2-4 m_Z^2\right){}^2} \Bigg\{ \left(Q^2-2 m_Z^2\right) \big(m_Z^2 \left(-m_i \left(6 m_j^2+5 Q^2\right)+6 m_i^2
   m_j+6 m_i^3+3 m_j \left(Q^2-2 m_j^2\right)\right)\nonumber\\
   &+2 Q^4 m_i\big)B_{ii}(m_Z^2)
   +Q^2  \left(m_i+m_j\right) \left(-3 Q^2 m_Z^2+2 m_Z^4+Q^4\right)\Big[B_{ij}(m_Z^2) -2  B_{ij}(Q^2) \Big]\nonumber\\
   & - \left(Q^2-2 m_Z^2\right) \big(2 Q^2 m_Z^4 \left(m_i+m_j\right)+m_Z^2 \big(2 m_i
   m_j^2 \left(Q^2-6 m_i^2\right)-6 m_j^3 \left(2 m_i^2+Q^2\right)\nonumber\\
   &+m_j \left(-2 Q^2
   m_i^2+6 m_i^4+Q^4\right)+6 m_i m_j^4-3 Q^4 m_i-10 Q^2 m_i^3+6 m_i^5+6
   m_j^5\big)\nonumber\\
   &+Q^4 m_i \left(2 \left(2 m_i-m_j\right)
   \left(m_i+m_j\right)+Q^2\right)\big) C_{iij}(Q^2)\nonumber\\
   &- Q^2 g_A \left(m_i+m_j\right) \left(-6 Q^2 m_Z^2+8 m_Z^4+Q^4\right)\Bigg\}+(i \leftrightarrow j),
\end{align}

\begin{align}
\label{}
 R^H_4(Q^2,m_i,m_j)= &\frac{1}{Q^4\left(Q^2-4 m_Z^2\right){}^2} \Bigg\{m_Z^2 \big(-6 m_i^2 m_j \left(Q^2-2 m_Z^2\right)-m_i \big(6 m_j^2 \left(Q^2-2
   m_Z^2\right)-10 Q^2 m_Z^2+Q^4\big)\nonumber\\
   &+6 m_i^3 \left(Q^2-2 m_Z^2\right)-3 m_j
   \left(Q^2-2 m_j^2\right) \left(Q^2-2 m_Z^2\right)\big)B_{ii}(m_Z^2)\nonumber\\
   &+ Q^2 g_V m_Z^2 \left(m_i-m_j\right) \left(2 m_Z^2+Q^2\right)\Big[B_{ij}(m_Z^2) -2  B_{ij}(Q^2) \Big]\nonumber\\
   & +\big(6 m_i^4 m_j m_Z^2 \left(Q^2-2 m_Z^2\right)-2 m_i^3 \big(Q^6-m_Z^2 \left(6
   m_j^2+5 Q^2\right) \left(Q^2-2 m_Z^2\right)\big)\nonumber\\
   &+2 m_i^2 m_j \left(Q^2-2
   m_Z^2\right) \left(Q^4-m_Z^2 \left(6 m_j^2+Q^2\right)\right)+m_i \big(4 m_Z^4
   \left(Q^2 m_j^2+3 m_j^4+2 Q^4\right)\nonumber\\
   &+m_Z^2 \left(2 Q^4 m_j^2-6 Q^2 m_j^4-7
   Q^6\right)+4 Q^2 m_Z^6+Q^8\big)-6 m_i^5 m_Z^2 \left(Q^2-2 m_Z^2\right)\nonumber\\
   &+m_j m_Z^2
   \left(Q^2-2 m_Z^2\right) \left(-6 Q^2 m_j^2+6 m_j^4+2 Q^2 m_Z^2+Q^4\right)\big)C_{iij}(Q^2)\nonumber\\
   &- Q^2  \left(m_i-m_j\right) \left(-6 Q^2 m_Z^2+8 m_Z^4+Q^4\right)\Bigg\}-(i \leftrightarrow j),
\end{align}

\begin{align}
\label{}
 R^Z_1(Q^2,m_i,m_j)=&\frac{1}{
   \left((m_Z+m_H)^2-P^2\right){}^2 \left((m_Z-m_H)^2-P^2\right){}^2}\Bigg\{ \frac{P^2}{2}   \Big[2 m_Z^2 \big(m_i \left(4 \left(m_H^2+P^2\right)+3
   m_j^2\right)\nonumber\\
   &-3 m_i^2 m_j-3 m_i^3+3 m_j^3\big)-\left(P^2-m_H^2\right) \big(m_i
   \left(-m_H^2-6 m_j^2+P^2\right)-3 m_j \left(-m_H^2+2 m_j^2+P^2\right)\nonumber\\
   &+6 m_i^2 m_j+6
   m_i^3\big)+m_Z^4 \left(-\left(7 m_i+3 m_j\right)\right)\Big]B_{ii}(P^2)+\frac{m_Z^2}{2}   \Big[2 m_Z^2 \big(-3 m_j \left(m_H^2+m_i^2\right)\nonumber\\
   &+m_i
   \left(m_H^2-3 m_i^2+4 P^2\right)+3 m_i m_j^2+3 m_j^3\big)-\left(P^2-m_H^2\right)
   \big(m_i \left(-m_H^2-6 m_j^2+7 P^2\right)\nonumber\\
   &+3 m_j \left(m_H^2-2 m_j^2+P^2\right)+6
   m_i^2 m_j+6 m_i^3\big)+m_Z^4 \left(-\left(m_i-3 m_j\right)\right)\Big]B_{ii}(m_Z^2)\nonumber\\
   &+ \frac{P^2}{2} \left(m_i+m_j\right) \left(4 m_Z^2
   \left(m_H^2+P^2\right)+\left(P^2-m_H^2\right){}^2-5 m_Z^4\right)B_{ij}(P^2)\nonumber\\
   &+ \frac{m_Z^2}{2} \left(m_i+m_j\right) \left(m_H^2 \left(4 P^2-2 m_Z^2\right)+m_H^4+4 P^2
   m_Z^2+m_Z^4-5 P^4\right)B_{ij}(m_Z^2)\nonumber\\
   &-  \left(m_i+m_j\right) \left(m_H^4 \left(m_Z^2+P^2\right)-2 m_H^2 \left(-4
   P^2 m_Z^2+m_Z^4+P^4\right)+\left(P^2-m_Z^2\right){}^2 \left(m_Z^2+P^2\right)\right)B_{ij}(m_H^2)\nonumber\\
 &+\frac{1}{4}  \Big[-m_H^6 \left(2 m_i^2 m_j+m_i \left(4 m_Z^2+3 P^2\right)+2
   m_i^3+P^2 m_j\right)+m_H^4 \big(m_Z^2 \big(6 m_i^2 m_j-P^2 m_i+6 m_i^3\nonumber\\
   &-3 P^2
   m_j\big)+P^2 \left(m_i+m_j\right) \left(-4 m_i m_j+4 m_i^2+6 m_j^2+3 P^2\right)+6
   m_i m_Z^4\big)\nonumber\\
   &+m_H^2 \big(m_Z^4 \left(-6 m_i^2 m_j+5 P^2 m_i-6 m_i^3+3 P^2
   m_j\right)+4 P^2 m_Z^2 \left(m_i+m_j\right) \left(2 m_i m_j-3 m_i^2+P^2\right)\nonumber\\
   &-P^2
   \big(2 m_i^3 \left(P^2-6 m_j^2\right)-6 m_i^2 m_j \left(2 m_j^2+P^2\right)+m_i
   \left(4 P^2 m_j^2+6 m_j^4+P^4\right)+6 m_i^4 m_j+6 m_i^5\nonumber\\
   &+3 m_j \left(4 P^2 m_j^2+2
   m_j^4+P^4\right)\big)-4 m_i m_Z^6\big)+m_H^8 m_i+m_Z^6 \left(2 m_i^2 m_j-P^2
   m_i+2 m_i^3+P^2 m_j\right)\nonumber\\
   &-P^2 m_Z^4 \left(m_i+m_j\right) \left(4 m_i m_j-8 m_i^2+6
   m_j^2+P^2\right)+P^2 m_Z^2 \big(-2 m_i^3 \left(6 m_j^2+5 P^2\right)\nonumber\\
   &-2 m_i^2 m_j
   \left(6 m_j^2+P^2\right)+m_i \left(8 P^2 m_j^2+6 m_j^4+P^4\right)+6 m_i^4 m_j+6
   m_i^5-P^4 m_j+6 m_j^5\big)\nonumber\\
   &+P^4 \left(2 \left(m_i+m_j\right) \left(-2 P^2 m_i m_j+3
   m_j^2 \left(P^2-2 m_i^2\right)+3 m_i^4+3 m_j^4\right)+P^4 m_j\right)+m_i
   m_Z^8\Big]C_{iji}(P^2)\nonumber\\
   &+\frac{1}{4}  \Big[ m_Z^6 \big(-m_j \left(3 m_H^2+4 m_i^2+P^2\right)+m_i
   \left(P^2-m_H^2\right)+2 m_i m_j^2+6 m_j^3\big)\nonumber\\
   &+m_i \left(P^2-m_H^2\right){}^3
   \left(-m_H^2+2 m_i \left(m_i+m_j\right)+P^2\right)+m_Z^4 \left(m_i+m_j\right)
   \big(2 m_H^2 \big(4 m_i m_j-m_i^2\nonumber\\
   &-6 m_j^2+2 P^2\big)+3 m_H^4+8 P^2 m_i m_j-2
   m_i^2 \left(6 m_j^2+5 P^2\right)+6 m_i^4+6 m_j^4-P^4\big)\nonumber\\
   &-m_Z^2
   \left(P^2-m_H^2\right) \big(6 m_j^3 \left(m_H^2+2 m_i^2+P^2\right)+2 m_i m_j^2
   \left(m_H^2+6 m_i^2+5 P^2\right)-m_j \big(4 P^2 m_H^2\nonumber\\
   &+m_H^4+4 P^2 m_i^2+6
   m_i^4+P^4\big)+m_i \left(4 m_H^2 \left(m_i^2-P^2\right)-3 m_H^4-8 P^2 m_i^2-6
   m_i^4+P^4\right)-6 m_i m_j^4\nonumber\\
   &-6 m_j^5\big)+m_j m_Z^8\Big]C_{jii}(P^2)\nonumber\\
   &+ \left(m_i+m_j\right) \left((m_H-m_Z)^2-P^2\right)
 \left((m_H+m_Z)^2-P^2\right)\left(-m_H^2+m_Z^2+P^2\right)
   \Bigg\}+(i\leftrightarrow j)
\end{align}

\begin{align}
\label{}
 R^Z_2(Q^2,m_i,m_j)=&\frac{1}{
   \left((m_Z+m_H)^2-P^2\right){}^2 \left((m_Z-m_H)^2-P^2\right){}^2}\Bigg\{ \frac{1}{2}  \left(-m_H^2+m_Z^2+P^2\right) \big(m_i \big(m_H^2 \left(4
   m_Z^2+P^2\right)\nonumber\\
   &-2 m_H^4+6 P^2 m_j^2+P^2 m_Z^2-2 m_Z^4+P^4\big)+3 P^2 m_j
   \left(m_H^2-2 m_j^2+m_Z^2-P^2\right)+6 P^2 m_i^2 m_j\nonumber\\
   &-6 P^2 m_i^3\big)B_{ii}(P^2)-\frac{1}{2} \left(-m_H^2+m_Z^2+P^2\right) \big(-m_Z^2 \big(m_i
   \left(m_H^2+6 m_j^2+P^2\right)+3 m_j \big(m_H^2-2 m_j^2\nonumber\\
   &+P^2\big)+6 m_i^2 m_j-6
   m_i^3\big)+2 m_i \left(P^2-m_H^2\right){}^2+m_Z^4 \left(-\left(m_i-3
   m_j\right)\right)\big)B_{ii}(m_Z^2)\nonumber\\
   &
   + \frac{\left(m_i-m_j\right)}{2} \left(-m_H^2+m_Z^2+P^2\right) \left(-m_H^2 \left(P^2-2
   m_Z^2\right)-m_H^4-m_Z^2 \left(m_Z^2+P^2\right)+2 P^4\right)B_{ij}(P^2)\nonumber\\
   &- \frac{\left(m_i-m_j\right)}{2} \left(-m_Z^4 \left(P^2-3 m_H^2\right)+2 P^2 m_Z^2
   \left(P^2-m_H^2\right)+\left(P^2-m_H^2\right){}^3-2 m_Z^6\right)B_{ij}(m_Z^2)\nonumber\\
   & - \left(m_i-m_j\right) \left(-3 m_H^4 \left(m_Z^2+P^2\right)+4 P^2 m_H^2
   m_Z^2+2 m_H^6+\left(P^2-m_Z^2\right){}^2 \left(m_Z^2+P^2\right)\right)B_{ij}(m_H^2)\nonumber\\
   &-\frac{1}{4}  \left(-m_H^2+m_Z^2+P^2\right) \big(m_H^4 \left(\left(2
   m_i+m_j\right) \left(2 m_i \left(m_j-m_i\right)+P^2\right)+m_i m_Z^2\right)\nonumber\\
   &+m_H^2
   \big(2 m_i^2 m_j \left(P^2-2 m_Z^2\right)+m_i \left(-2 m_Z^2 \left(2
   m_j^2+P^2\right)+2 P^2 m_j^2+m_Z^4-P^4\right)\nonumber\\
   &+2 m_i^3 \left(4 m_Z^2+P^2\right)-2 P^2
   m_j \left(3 m_j^2-2 m_Z^2+P^2\right)\big)+m_H^6 \left(-m_i\right)\nonumber\\
   &+m_Z^4 \left(2
   m_i+m_j\right) \left(2 m_i \left(m_j-m_i\right)+P^2\right)+P^2 m_Z^2 \big(-m_i
   \left(P^2-2 m_j^2\right)+2 m_i^2 m_j+2 m_i^3\nonumber\\
   &-2 m_j \left(3
   m_j^2+P^2\right)\big)+P^2 \big(6 m_j^3 \left(P^2-2 m_i^2\right)-4 m_i m_j^2
   \left(P^2-3 m_i^2\right)+m_j \left(-4 P^2 m_i^2+6 m_i^4+P^4\right)\nonumber\\
   &-6 m_i m_j^4+2
   m_i^3 \left(P^2-3 m_i^2\right)+6 m_j^5\big)-m_i m_Z^6\big)C_{iji}(P^2)\nonumber\\
   &+\frac{1}{4}  \left(-m_H^2+m_Z^2+P^2\right) \big(m_Z^4 \left(m_i \left(m_H^2+4
   m_j^2+P^2\right)+2 m_j \left(m_H^2-3 m_j^2+P^2\right)+4 m_i^2 m_j-2 m_i^3\right)\nonumber\\
   &+m_i
   \left(P^2-m_H^2\right){}^2 \left(m_H^2+2 \left(m_i-m_j\right) \left(2
   m_i+m_j\right)+P^2\right)-m_Z^2 \big(-6 m_j^3 \left(m_H^2+2 m_i^2+P^2\right)\nonumber\\
   &+2 m_i
   m_j^2 \left(m_H^2+6 m_i^2+P^2\right)+m_j \left(2 m_i^2 \left(m_H^2+P^2\right)+4 P^2
   m_H^2+m_H^4+6 m_i^4+P^4\right)\nonumber\\
   &+2 m_i \left(m_i^2 \left(m_H^2+P^2\right)-P^2
   m_H^2+m_H^4-3 m_i^4+P^4\right)-6 m_i m_j^4+6 m_j^5\big)-m_j m_Z^6\big)C_{jii}(P^2)\nonumber\\
   &+ \left(m_i-m_j\right)  \left((m_H-m_Z)^2-P^2\right)
 \left((m_H+m_Z)^2-P^2\right) \left(-m_H^2+m_Z^2+P^2\right)\Bigg\}-(i\leftrightarrow j)
\end{align}

\begin{align}
\label{}
 R^Z_3(Q^2,m_i,m_j)=&\frac{1}{
   \left((m_Z+m_H)^2-P^2\right){}^2 \left((m_Z-m_H)^2-P^2\right){}^2}\Bigg\{ \frac{1}{2}  \left(-m_H^2+m_Z^2+P^2\right) \big(m_i \big(m_H^2 \left(4
   m_Z^2+P^2\right)\nonumber\\
   &-2 m_H^4+6 P^2 m_j^2+P^2 m_Z^2-2 m_Z^4+P^4\big)+3 P^2 m_j
   \left(-m_H^2+2 m_j^2-m_Z^2+P^2\right)-6 P^2 m_i^2 m_j\nonumber\\
   &-6 P^2 m_i^3\big)B_{ii}(P^2)-\frac{1}{2} \left(-m_H^2+m_Z^2+P^2\right) \big(-m_Z^2 \big(-3 m_j
   \left(m_H^2+2 m_i^2+P^2\right)\nonumber\\
   &+m_i \left(m_H^2-6 m_i^2+P^2\right)+6 m_i m_j^2+6
   m_j^3\big)+2 m_i \left(P^2-m_H^2\right){}^2+m_Z^4 \left(-\left(m_i+3
   m_j\right)\right)\big)B_{ii}(m_Z^2)\nonumber\\
   &+\frac{1}{2}\left(m_i+m_j\right) \left(-m_H^2+m_Z^2+P^2\right) \left(-m_H^2
   \left(P^2-2 m_Z^2\right)-m_H^4-m_Z^2 \left(m_Z^2+P^2\right)+2 P^4\right)B_{ij}(P^2)\nonumber\\
   &+\frac{1}{2} \left(m_i+m_j\right) \left(m_Z^4 \left(P^2-3 m_H^2\right)+2 P^2
   m_Z^2 \left(m_H^2-P^2\right)-\left(P^2-m_H^2\right){}^3+2 m_Z^6\right)B_{ij}(m_Z^2)\nonumber\\
   &- \left(m_i+m_j\right) \left(-3 m_H^4 \left(m_Z^2+P^2\right)+4 P^2 m_H^2
   m_Z^2+2 m_H^6+\left(P^2-m_Z^2\right){}^2 \left(m_Z^2+P^2\right)\right)B_{ij}(m_H^2)\nonumber\\
   & +\frac{1}{4} \left(-m_H^2+m_Z^2+P^2\right) \big(m_H^4 \left(\left(2
   m_i-m_j\right) \left(2 m_i \left(m_i+m_j\right)-P^2\right)-m_i m_Z^2\right)\nonumber\\
   & +m_H^2
   \big(2 m_i^2 m_j \left(P^2-2 m_Z^2\right)+m_i \left(2 m_Z^2 \left(2
   m_j^2+P^2\right)-2 P^2 m_j^2-m_Z^4+P^4\right)-2 m_i^3 \left(4 m_Z^2+P^2\right)\nonumber\\
   & -2 P^2
   m_j \left(3 m_j^2-2 m_Z^2+P^2\right)\big)+m_H^6 m_i+m_Z^4 \left(2 m_i-m_j\right)
   \left(2 m_i \left(m_i+m_j\right)-P^2\right)\nonumber\\
   & +P^2 m_Z^2 \left(m_i \left(P^2-2
   m_j^2\right)+2 m_i^2 m_j-2 m_i^3-2 m_j \left(3 m_j^2+P^2\right)\right)+P^2 \big(4
   m_i m_j^2 \left(P^2-3 m_i^2\right)\nonumber\\
   & +6 m_j^3 \left(P^2-2 m_i^2\right)+m_j \left(-4 P^2
   m_i^2+6 m_i^4+P^4\right)+6 m_i m_j^4-2 P^2 m_i^3+6 m_i^5+6 m_j^5\big)\nonumber\\
   & +m_i
   m_Z^6\big)C_{iji}(P^2)+\frac{1}{4}  \left(-m_H^2+m_Z^2+P^2\right) \big(m_Z^4 \big(-2 m_j
   \left(m_H^2+2 m_i^2+P^2\right)\nonumber\\
   &+m_i \left(m_H^2-2 m_i^2+P^2\right)+4 m_i m_j^2+6
   m_j^3\big)+m_i \left(P^2-m_H^2\right){}^2 \big(m_H^2+2 \left(2 m_i-m_j\right)
   \left(m_i+m_j\right)\nonumber\\
   &+P^2\big)+m_Z^2 \big(-2 m_i^3 \left(m_H^2+6
   m_j^2+P^2\right)+2 m_i^2 m_j \left(m_H^2-6 m_j^2+P^2\right)-2 m_i \big(m_j^2
   \left(m_H^2+P^2\right)\nonumber\\
   &-P^2 m_H^2+m_H^4-3 m_j^4+P^4\big)+m_j \left(-6 m_j^2
   \left(m_H^2+P^2\right)+4 P^2 m_H^2+m_H^4+6 m_j^4+P^4\right)\nonumber\\
   &+6 m_i^4 m_j+6
   m_i^5\big)+m_j m_Z^6\big)C_{jii}(P^2)\nonumber\\
   &+\left(m_i+m_j\right) \left((m_H-m_Z)^2-P^2\right)
 \left((m_H+m_Z)^2-P^2\right) \left(-m_H^2+m_Z^2+P^2\right)
\Bigg\}   +(i\leftrightarrow j)
\end{align}

\begin{align}
\label{}
 R^Z_4(Q^2,m_i,m_j)=&\frac{1}{
   \left((m_Z+m_H)^2-P^2\right){}^2 \left((m_Z-m_H)^2-P^2\right){}^2}\Bigg\{\frac{1}{2}  P^2 \big(2 m_Z^2 \big(m_i \left(4 \left(m_H^2+P^2\right)+3
   m_j^2\right)+3 m_i^2 m_j\nonumber\\
   &-3 m_i^3-3 m_j^3\big)-\left(P^2-m_H^2\right) \big(m_i
   \left(-m_H^2-6 m_j^2+P^2\right)+3 m_j \left(-m_H^2+2 m_j^2+P^2\right)-6 m_i^2 m_j\nonumber\\
   &+6
   m_i^3\big)+m_Z^4 \left(3 m_j-7 m_i\right)\big)B_{ii}(P^2)+\frac{1}{2} m_Z^2 \Big[2 m_Z^2 \big(3 m_j \left(m_H^2+m_i^2\right)+m_i
   \left(m_H^2-3 m_i^2+4 P^2\right)\nonumber\\
   &+3 m_i m_j^2-3 m_j^3\big)-\left(P^2-m_H^2\right)
   \big(-3 m_j \left(m_H^2+2 m_i^2+P^2\right)+m_i \left(-m_H^2+6 m_i^2+7 P^2\right)\nonumber\\
   &-6
   m_i m_j^2+6 m_j^3\big)+m_Z^4 \left(-\left(m_i+3 m_j\right)\right)\Big]B_{ii}(m_Z^2)\nonumber\\
   &+\frac{1}{2}P^2 \left(m_i-m_j\right) \left(4 m_Z^2
   \left(m_H^2+P^2\right)+\left(P^2-m_H^2\right){}^2-5 m_Z^4\right)B_{ij}(P^2)\nonumber\\
   &+\frac{1}{2}  m_Z^2 \left(m_i-m_j\right) \left(m_H^2 \left(4 P^2-2
   m_Z^2\right)+m_H^4+4 P^2 m_Z^2+m_Z^4-5 P^4\right)B_{ij}(m_Z^2)\nonumber\\
   &-\left(m_i-m_j\right) \left(m_H^4 \left(m_Z^2+P^2\right)-2 m_H^2 \left(-4 P^2
   m_Z^2+m_Z^4+P^4\right)+\left(P^2-m_Z^2\right){}^2 \left(m_Z^2+P^2\right)\right)B_{ij}(m_H^2)	\nonumber\\
   &+\frac{1}{4}  \Big[m_H^6 \left(2 m_i^2 m_j-m_i \left(4 m_Z^2+3 P^2\right)-2
   m_i^3+P^2 m_j\right)+m_H^4 \big(m_Z^2 \big(-6 m_i^2 m_j-P^2 m_i\nonumber\\
   &+6 m_i^3+3 P^2
   m_j\big)+P^2 \left(m_i-m_j\right) \left(4 m_i m_j+4 m_i^2+6 m_j^2+3 P^2\right)+6
   m_i m_Z^4\big)\nonumber\\
   &+m_H^2 \big(m_Z^4 \left(6 m_i^2 m_j+5 P^2 m_i-6 m_i^3-3 P^2
   m_j\right)+4 P^2 m_Z^2 \left(m_i-m_j\right) \left(-2 m_i m_j-3 m_i^2+P^2\right)\nonumber\\
   &-P^2
   \big(2 m_i^3 \left(P^2-6 m_j^2\right)+6 m_i^2 m_j \left(2 m_j^2+P^2\right)+m_i
   \left(4 P^2 m_j^2+6 m_j^4+P^4\right)-6 m_i^4 m_j+6 m_i^5\nonumber\\
   &-3 m_j \left(4 P^2 m_j^2+2
   m_j^4+P^4\right)\big)-4 m_i m_Z^6\big)+m_H^8 m_i-m_Z^6 \left(m_j \left(2
   m_i^2+P^2\right)+m_i \left(P^2-2 m_i^2\right)\right)\nonumber\\
   &-P^2 m_Z^4 \left(m_i-m_j\right)
   \left(-4 m_i m_j-8 m_i^2+6 m_j^2+P^2\right)+P^2 m_Z^2 \big(-2 m_i^3 \left(6 m_j^2+5
   P^2\right)\nonumber\\
   &+2 m_i^2 m_j \left(6 m_j^2+P^2\right)+m_i \left(8 P^2 m_j^2+6
   m_j^4+P^4\right)-6 m_i^4 m_j+6 m_i^5+m_j \left(P^4-6 m_j^4\right)\big)\nonumber\\
   &+P^4 \left(2
   \left(m_i-m_j\right) \left(2 P^2 m_i m_j+3 m_j^2 \left(P^2-2 m_i^2\right)+3 m_i^4+3
   m_j^4\right)-P^4 m_j\right)+m_i m_Z^8\Big]C_{iji}(P^2)\nonumber\\
   &+\frac{1}{4} \Big[m_Z^6 \left(m_i \left(-m_H^2+2 m_j^2+P^2\right)+m_j \left(3
   m_H^2-6 m_j^2+P^2\right)+4 m_i^2 m_j\right)\nonumber\\
   &+m_i \left(P^2-m_H^2\right){}^3
   \left(-m_H^2+2 m_i \left(m_i-m_j\right)+P^2\right)-m_Z^4 \left(m_i-m_j\right)
   \big(2 m_H^2 \big(4 m_i m_j+m_i^2+6 m_j^2\nonumber\\
   &-2 P^2\big)-3 m_H^4+8 P^2 m_i m_j+2
   m_i^2 \left(6 m_j^2+5 P^2\right)-6 m_i^4-6 m_j^4+P^4\big)\nonumber\\
   &-m_Z^2
   \left(P^2-m_H^2\right) \big(-6 m_j^3 \left(m_H^2+2 m_i^2+P^2\right)+2 m_i m_j^2
   \left(m_H^2+6 m_i^2+5 P^2\right)\nonumber\\
   &+m_j \left(4 P^2 m_H^2+m_H^4+4 P^2 m_i^2+6
   m_i^4+P^4\right)+m_i \big(4 m_H^2 \left(m_i^2-P^2\right)-3 m_H^4-8 P^2 m_i^2\nonumber\\
   &-6
   m_i^4+P^4\big)-6 m_i m_j^4+6 m_j^5\big)-m_j m_Z^8\Big]C_{jii}(P^2)\nonumber\\
   & \left(m_i-m_j\right)  \left((m_H-m_Z)^2-P^2\right)
 \left((m_H+m_Z)^2-P^2\right) \left(-m_H^2+m_Z^2+P^2\right)
\Bigg\}    -(i\leftrightarrow j)
\end{align}

The functions $T^V_{1,2,3,4}$ ($V=H$, $Z$) for contributions of Type II are given as follows:

\begin{align}
\label{}
 T^H_1(Q^2,m_i,m_j)= &\frac{1}{Q^2\left(Q^2-4 m_Z^2\right)}\Bigg\{2  \left(m_Z^2 \left(m_i+m_j\right)-Q^2 m_i\right)B_{ii}(m_Z^2)+2 Q^2  \left(m_j-m_i\right)\Big[B_{ij}(m_Z^2) -2  B_{ij}(Q^2) \Big]\nonumber\\,
 & \left(2 m_i^2-2 m_j^2+Q^2\right) \left(Q^2 m_i-m_Z^2 \left(m_i+m_j\right)\right)C_{iij}(Q^2)\Bigg\}-(i \leftrightarrow j),
\end{align}

\begin{align}
\label{}
 T^H_2(Q^2,m_i,m_j)=&\frac{1}{Q^2\left(Q^2-4 m_Z^2\right)}\Bigg\{2  m_Z^2 \left(m_i-m_j\right)B_{ii}(m_Z^2)+ \big[m_Z^2 \big(m_i \left(2 m_j^2+3 Q^2\right)+2 m_i^2 m_j-2 m_i^3\nonumber\\
   &+m_j
   \left(Q^2-2 m_j^2\right)\big)-Q^4 m_i\big]C_{iij}(Q^2)\Big\} +(i \leftrightarrow j),
\end{align}

\begin{align}
\label{}
 T^H_3(Q^2,m_i,m_j)=&\frac{1}{Q^2\left(Q^2-4 m_Z^2\right)}\Bigg\{2 m_Z^2 \left(m_i+m_j\right)B_{ii}(m_Z^2)+ \big[m_Z^2 \big(-m_j \left(2 m_i^2+Q^2\right)+2 m_i m_j^2+3 Q^2 m_i\nonumber\\
   &-2 m_i^3+2
   m_j^3\big)-Q^4 m_i\big] C_{iij}(Q^2)\Bigg\}-(i \leftrightarrow j),
\end{align}

\begin{align}
\label{}
 T^H_4(Q^2,m_i,m_j)=&\frac{1}{Q^2\left(Q^2-4 m_Z^2\right)}\Bigg\{-2 \left(m_Z^2 \left(m_j-m_i\right)+Q^2 m_i\right)B_{ii}(m_Z^2)-2 Q^2  \left(m_i+m_j\right)\Big[B_{ij}(m_Z^2) -2  B_{ij}(Q^2) \Big]\nonumber\\,
 &+ \left(2 m_i^2-2 m_j^2+Q^2\right) \left(m_Z^2 \left(m_j-m_i\right)+Q^2 m_i\right)C_{iij}(Q^2)\Bigg\}+(i \leftrightarrow j) ,
\end{align}

\begin{align}
\label{}
 T^Z_1(Q^2,m_i,m_j)= &\frac{1}{((m_H-m_Z)^2-P^2)((m_H+m_Z)-P^2)}\Bigg\{\left(m_i \left(m_Z^2-m_H^2\right)+P^2 m_j\right)B_{ii}(P^2)\nonumber\\
 & +\left(m_i \left(P^2-m_H^2\right)+m_j m_Z^2\right)B_{ii}(m_Z^2)-\frac{1}{2} \left(m_i-m_j\right) \left(m_H^2-m_Z^2+P^2\right)B_{ij}(P^2)\nonumber\\
 &+\frac{1}{2} \left(m_i-m_j\right) \left(-m_H^2-m_Z^2+P^2\right)B_{ij}(m_Z^2)+2 m_H^2 \left(m_i-m_j\right)B_{ij}(m_H^2)\nonumber\\
 &+\frac{1}{4}  \left(-m_H^2-2 m_i^2+2 m_j^2-m_Z^2+P^2\right) \left(m_i
   \left(m_Z^2-m_H^2\right)+P^2 m_j\right)C_{iji}(P^2)\nonumber\\
   &-\frac{1}{4} \left(m_H^2+2 m_i^2-2 m_j^2-m_Z^2+P^2\right) \left(m_i
   \left(P^2-m_H^2\right)+m_j m_Z^2\right)C_{jii}
  \Bigg\}-(i \leftrightarrow j),
\end{align}

\begin{align}
\label{}
 T^Z_2(Q^2,m_i,m_j)= &\frac{1}{((m_H-m_Z)^2-P^2)((m_H+m_Z)-P^2)}\Bigg\{P^2 \left(m_i-m_j\right)B_{ii}(P^2)+ m_Z^2 \left(m_i-m_j\right)B_{ii}(m_Z^2)\nonumber\\
 &+ \frac{1}{4}  \Big[m_H^2 \left(P^2 \left(m_i+m_j\right)+2 m_i
   m_Z^2\right)+m_H^4 \left(-m_i\right)+P^2 m_Z^2 \left(m_i+m_j\right)\nonumber\\
   &-P^2 \left(2
   \left(m_i-m_j\right){}^2 \left(m_i+m_j\right)+P^2 m_j\right)-m_i m_Z^4\Big]C_{iji}(P^2)\nonumber\\
   &+\frac{1}{4} \Big[m_Z^2 \left(m_i+m_j\right) \left(m_H^2-2
   \left(m_i-m_j\right){}^2+P^2\right)-m_i \left(P^2-m_H^2\right){}^2-m_j m_Z^4\Big]C_{jii}(P^2)
  \Bigg\}\nonumber\\&+(i \leftrightarrow j),
\end{align}

\begin{align}
\label{}
 T^Z_3(Q^2,m_i,m_j)= &\frac{1}{((m_H-m_Z)^2-P^2)((m_H+m_Z)-P^2)}\Bigg\{ P^2 \left(m_i+m_j\right)B_{ii}(P^2)+ m_Z^2 \left(m_i+m_j\right)B_{ii}(m_Z^2)\nonumber\\
 &+\frac{1}{4}  \Big[m_H^2 \left(m_i \left(2 m_Z^2+P^2\right)-P^2
   m_j\right)+m_H^4 \left(-m_i\right)+P^2 m_Z^2 \left(m_i-m_j\right)\nonumber\\
   &+P^2 \left(P^2
   m_j-2 \left(m_i-m_j\right) \left(m_i+m_j\right){}^2\right)-m_i m_Z^4\Big]C_{iji}(P^2)\nonumber\\
   &+\frac{1}{4}  \Big[m_Z^2 \left(m_i-m_j\right) \left(m_H^2-2
   \left(m_i+m_j\right){}^2+P^2\right)-m_i \left(P^2-m_H^2\right){}^2+m_j m_Z^4\Big]C_{jii}(P^2)
  \Bigg\}\nonumber\\&-(i \leftrightarrow j),
\end{align}

\begin{align}
\label{}
 T^Z_4(Q^2,m_i,m_j)= &\frac{1}{((m_H-m_Z)^2-P^2)((m_H+m_Z)-P^2)}\Bigg\{-\left(m_i \left(m_H^2-m_Z^2\right)+P^2 m_j\right)B_{ii}(P^2)\nonumber\\
 &+ \left(m_i \left(P^2-m_H^2\right)-m_j m_Z^2\right)B_{ii}(m_Z^2)-\frac{1}{2} \left(m_i+m_j\right) \left(m_H^2-m_Z^2+P^2\right)B_{ij}(P^2)\nonumber\\
 &+\frac{1}{2} \left(m_i+m_j\right) \left(-m_H^2-m_Z^2+P^2\right)B_{ij}(m_Z^2)+2 m_H^2 \left(m_i+m_j\right)B_{ij}(m_H^2)\nonumber\\
 & -\frac{1}{4}  \left(-m_H^2-2 m_i^2+2 m_j^2-m_Z^2+P^2\right) \left(m_i
   \left(m_H^2-m_Z^2\right)+P^2 m_j\right)C_{iji}(P^2)\nonumber\\
   &-\frac{1}{4} \text{ga} \left(m_H^2+2 m_i^2-2 m_j^2-m_Z^2+P^2\right) \left(m_i
   \left(P^2-m_H^2\right)-m_j m_Z^2\right)C_{jii}(P^2)
   \Bigg\}+(i \leftrightarrow j),
\end{align}

\bibliography{BiblioH}
\end{document}